\documentclass[lettersize,journal]{IEEEtran}
\usepackage{amsmath,amsfonts,amsthm}
\usepackage{algorithmic}
\usepackage{algorithm}
\usepackage{array}
\usepackage[caption=false,font=normalsize,labelfont=sf,textfont=sf]{subfig}
\usepackage{textcomp}
\usepackage{stfloats}
\usepackage{verbatim}
\usepackage{graphicx}
\usepackage{booktabs}
\usepackage{siunitx}
\usepackage{makecell}
\usepackage{caption}
\usepackage{enumitem}
\usepackage{multirow}
\usepackage{microtype}
\usepackage{hyperref}
\usepackage[margin=1in]{geometry}
\usepackage{tikz}
\usepackage{amsmath,amssymb}
\usepackage{xcolor}
\usepackage{array}
\usetikzlibrary{arrows.meta, positioning, fit, backgrounds, shapes.geometric, calc}

\definecolor{clientbg}{RGB}{220,235,255}   
\definecolor{serverbg}{RGB}{220,255,230}   
\definecolor{arrowcol}{RGB}{50,50,150}
\definecolor{boxborder}{RGB}{80,80,200}
\definecolor{stepbg}{RGB}{245,245,245}
\definecolor{critbg}{RGB}{255,240,200}     
\definecolor{headerbg}{RGB}{60,90,160}
\definecolor{headertext}{RGB}{255,255,255}
\usepackage{cite}
\usepackage{balance}
\allowdisplaybreaks
\newtheorem{theorem}{Theorem}
\newtheorem{definition}{Definition}
\begin{document}
\title{AmphiKey: A Dual-Mode Secure Authenticated Key Encapsulation Protocol for Smart Grid}
\author{
    Kazi Hassan Shakib, \textit{Graduate Student Member, IEEE},
    Muhammad Asfand Hafeez, \textit{Graduate Student Member, IEEE},
    and Arslan Munir, \textit{Senior~Member,~IEEE}
    \thanks{Kazi Hassan Shakib is with the Department of Computer Science, Kansas State University, Manhattan, Kansas, 66506, USA (e-mail: kshakib@ksu.edu).}
    \thanks{Muhammad Asfand Hafeez and Arslan Munir are with the Department of Electrical Engineering and Computer Science, Florida Atlantic University, Boca Raton, FL 33431 USA (e-mail: mhafeez2024@fau.edu; arslanm@fau.edu).}
}
\maketitle
\begin{abstract}
AmphiKey is a dual-mode post-quantum/ traditional (PQ/T) hybrid authenticated key encapsulation mechanism (AKEM) designed to secure smart grid communications against classical and quantum threats, as well as physical side-channel attacks (SCAs). It offers two operational modes: an \textbf{Authenticated Mode} and a \textbf{Deniable Mode}. The Authenticated Mode combines ephemeral ML-KEM-768 and X25519 (instantiated as IND-CCA2 KEMs using HPKE-style DHKEM binding) with long-term Raccoon DSA, achieving \textbf{`OR'} confidentiality and \textbf{`AND'} authenticity, making the shared secret secure if either KEM holds, and authenticity requires both a verified EUF-CMA signature and successful decapsulation. The masking-friendly Raccoon DSA provides SCA resistance for the signing key; the scope and residual limitations of this protection are explicitly characterized. The Deniable Mode replaces the signature with a lightweight HMAC-based tag derived solely from ephemeral KEM secrets, enabling formal sender deniability with third-party transcript indistinguishability (without receiver-side entity authentication of the sender) at a 93\% lower computational cost. A cryptographically bound \texttt{MODE} flag prevents downgrade attacks; per-session ephemeral keys and nonce binding provide replay resistance. Complete formal security proofs for both modes are presented using a sequence-of-games methodology. We evaluate AmphiKey on a heterogeneous testbed (AMD Ryzen 5 server, Raspberry Pi 500 client). The Deniable Mode completes handshakes in 0.15\,ms (server) and 0.41\,ms (Pi). The Authenticated Mode requires 4.8\,ms (Pi signing) and 0.84\,ms (server verification), with a 12,644-byte payload suitable for auditable command-and-control traffic.
\end{abstract}
\section{Introduction}
\IEEEPARstart{S}{mart} grids grids represent a transition from traditional power
grids, integrating advanced digital technologies  into power infrastructure, expanding the attack surface and making cybersecurity considerations imperative. Communication networks carry sensitive operational commands and customer data; compromise can cause widespread outages, financial loss, and safety risks to critical infrastructure~\cite{faquir2021cybersecurity}. A  additional threat is the development of large-scale quantum computers capable of breaking RSA and Elliptic Curve Cryptography (ECC) via Shor's algorithm, which solves integer factorization and the discrete logarithm problem in polynomial time~\cite{nistpqc,AHMAD2025126118}. Adversaries already harvest encrypted traffic today for decryption, and once quantum computers mature, this becomes a ``harvest now, decrypt later'' threat that is particularly acute for long-lived grid infrastructure whose operational lifetime may span 20--30 years~\cite{bindel2019hybrid}.

Lattice-based problems such as Module Learning With Errors (MLWE) resist all known quantum algorithms beyond Grover's quadratic speedup~\cite{nistfips203}, underpinning NIST's standardized post-quantum cryptography (PQC) suite finalized in 2024. The case for \emph{hybrid} (PQ/classical) protocols is compelling: they defend simultaneously against (1) the eventual quantum threat to classical cryptography, and (2) unforeseen vulnerabilities in newly standardized PQC algorithms.
Smart grid devices also face \emph{physical} threats distinct from network-layer attacks. Components such as smart meters, protective relays, and Phasor Measurement Units (PMUs) are often field-deployed in physically accessible locations, making them susceptible to SCAs that extract cryptographic keys via power or electromagnetic (EM) analysis~\cite{chaudhry2017lightweight,renitaSCA}. Masking countermeasures for standard lattice schemes like Dilithium impose $>200\times$ overhead when subsequently done as post-hoc~\cite{Barthemasking} which makes them infeasible on constrained devices and specialized first-order masking gadgets for Dilithium still impose $40$--$60\times$ overhead~\cite{Coron2023Gadgets}. This creates a critical need for signature schemes designed with masking as an architectural priority from the outset.
The literature reveals a persistent gap: no single protocol simultaneously provides (a) hybrid PQ/classical confidentiality, (b) formal sender deniability as an alternative authentication mode, (c) SCA-resistant authentication via a masking-friendly signature scheme, and (d) explicit analysis of SCA scope tailored to the smart grid IoT deployment model. In response, we present \textbf{AmphiKey}, a novel dual-mode AKEM that addresses all four requirements in a single, formally analyzed framework.
The primary contributions are:
\begin{enumerate}[leftmargin=*]
    \item \textbf{Dual-Mode AKEM with Downgrade Protection}: A single protocol offering a non-repudiable \textbf{Authenticated Mode} and a privacy-preserving \textbf{Deniable Mode}, with a cryptographically bound \texttt{MODE} flag that prevents active downgrade attacks via HKDF key derivation divergence.
    \item \textbf{Authenticated Mode}: Combines ML-KEM-768 and X25519 (IND-CCA2 via HPKE DHKEM) with Raccoon DSA, achieving `OR' confidentiality and joint-confirmation `AND' authenticity with SCA-resistant signing. Non-repudiable by EUF-CMA.
    \item \textbf{Deniable Mode}: HMAC-based tag derived from ephemeral KEM secrets provides formal third-party sender deniability which is proven using a perfect simulation argument and ciphertext integrity under the receiver's public key. Deniable Mode does not provide receiver-side entity authentication of the sender; device identity is established at the session provisioning layer.
    \item \textbf{Complete Formal Security Proofs}: Full sequence-of-games reductions for both modes and sender deniability via simulator construction with corruption model analysis.
    \item \textbf{Explicit SCA Scope Characterization}: Formal identification of SCA-protected components (Raccoon signing) vs.\ components requiring deployment-level countermeasures (ML-KEM, X25519).
    \item \textbf{Comprehensive Evaluation}: Benchmarks on AMD Ryzen 5 and Raspberry Pi 500 covering key generation, handshake latency, throughput, payload sizes, and CPU cycle energy proxies.
\end{enumerate}
The paper is organized as follows. Section~\ref{sec:background} provides background. Section~\ref{sec:related} presents the related work and comparison. Section~\ref{sec:methodology} details the protocol. Section~\ref{sec:security} presents the formal security analysis. Section~\ref{sec:setup} describes the experimental setup. Section~\ref{sec:results} presents results. Section~\ref{sec:conclusion} concludes.
\vspace{-3mm}
\section{Background}
\label{sec:background}
\subsection{Smart Grid Threat Model}
Smart grid infrastructure is susceptible to three principal attack categories. \emph{Network-layer attacks} include eavesdropping on metering telemetry, man-in-the-middle injection of false SCADA commands, and denial-of-service flooding of substation controllers~\cite{bhusal2021cybersecurity}. \emph{Cryptanalytic attacks} include classical algebraic attacks (on RSA, ECC) and quantum attacks (Shor's algorithm) that threaten long-term confidentiality of archived traffic. \emph{Physical attacks} include differential power analysis (DPA), simple power analysis (SPA), and EM side-channel attacks against field-deployed cryptographic hardware with no tamper-proof enclosure~\cite{renitaSCA}. Any practical protocol for the smart grid must address all three layers simultaneously; no prior work does so within a unified dual-mode AKEM framework with explicit SCA scope analysis.
\vspace{-3mm}
\subsection{Security Primitives and Assumptions}
\textbf{ML-KEM-768} (NIST FIPS~203, 2024)~\cite{nistfips203} is a module-lattice KEM based on the Kyber construction. It is assumed IND-CCA2 secure under the MLWE hardness assumption over module lattices with dimension $k=3$ and modulus $q=3329$. Ciphertext size: 1088 bytes; public key: 1184 bytes; security level: NIST Level~3 ($\approx$AES-192 classical, $\approx$AES-128 quantum).\\
\textbf{X25519} provides a Diffie-Hellman function over Curve25519 (a Montgomery curve)~\cite{rfc7748}. Raw X25519 provides only IND-CPA security. In AmphiKey, X25519 is instantiated using \textbf{HPKE DHKEM(X25519, HKDF-SHA256)} (RFC~9180, KEM ID 0x0020)~\cite{rfc9180}. This construction computes
\[
\begin{aligned}
\mathrm{dh}
&= X25519\!\left(k_{\mathrm{pr}_{\mathrm{eph}}},\,
k_{\mathrm{pub}_{r}}\right), \\[4pt]
\mathrm{kem\_context}
&= k_{\mathrm{pub}_{\mathrm{eph}}} \,\|\, k_{\mathrm{pub}_{r}}, \\[4pt]
\mathrm{shared\_key}
&= \mathrm{HKDF\text{-}SHA256.LabeledExtract} \! \big(\\
    \text{``HPKE-v1''}, \mathrm{dh},
&\qquad \text{``shared\_secret''},\, \mathrm{kem\_context}
\big).
\end{aligned}
\]. This binding to the suite label, sender ephemeral public key, and receiver public key prevents multi-user and key-substitution attacks. Alwen et al.~\cite{alwen2021hpke} (Theorem~3) prove that this DHKEM instantiation achieves IND-CCA2 security under the gap-CDH assumption in the random oracle model. AmphiKey uses this suite precisely; we inherit IND-CCA2 of X25519-DHKEM directly from Alwen et al. Ciphertext size: 32 bytes; public key: 32 bytes.\\
\textbf{Raccoon DSA}~\cite{raccoon2017,delpino2024raccoon} is a lattice-based digital signature assumed EUF-CMA secure, architecturally designed for efficient high-order masking with $O(d \log d)$ overhead for $d$ shares, without rejection sampling. While Raccoon's baseline cost is higher than ML-DSA-65, this comparison is misleading for SCA-sensitive deployments: early masked Dilithium was $>200\times$ slower~\cite{Barthemasking}, and recent specialized gadgets still impose $40$--$60\times$ overhead for first-order security~\cite{Coron2023Gadgets}. Raccoon's performance metric inherently includes robust SCA protection.
\\
\textbf{ML-DSA-65} (NIST FIPS~204)~\cite{nistfips204} is the NIST-standardized lattice signature (derived from Dilithium3), included as a performance comparison baseline. AmphiKey’s Authenticated Mode can also use ML-DSA-65 (3,293-byte signatures) at the cost of requiring separate masking countermeasures.~\cite{Barthemasking,Coron2023Gadgets}. Signature size: 3,293 bytes; public key: 1,952 bytes.\\
\textbf{Ascon-128}~\cite{wu2016acorn} is the NIST-standardized lightweight AEAD (2023). It uses a 320-bit state with an SPN-based permutation, designed for 64-bit platforms and hardware-constrained devices alike. Negligible performance overhead ($<$3\,$\mu$s per encryption in our tests) makes it ideal for high-frequency smart grid payload encryption.\\
\textbf{HKDF-SHA-256}~\cite{rfc5869} is a two-phase KDF (Extract + Expand) based on HMAC-SHA-256. \\
\textbf{HMAC-SHA-256}~\cite{rfc2104} provides a pseudorandom function over arbitrary-length inputs. Both are assumed to behave as secure PRFs; their SHA-256 hash core resists quantum attacks beyond a quadratic speedup via Grover's algorithm (effective security: 128-bit post-quantum).
\vspace{-3mm}
\subsection{Formal Definition: IND-CCA2 Security for KEMs}
\label{subsec:indcca2_def}
\begin{definition}[IND-CCA2 KEM Security]
Let $\Pi = (\mathrm{Gen}, \mathrm{Enc}, \mathrm{Dec})$ be a KEM with security parameter $\lambda$. The \emph{IND-CCA2 game} is:
\begin{enumerate}[leftmargin=*]
    \item \textbf{Setup}: $(k_{\mathrm{pr}}, k_{\mathrm{pub}}) \leftarrow \mathrm{Gen}(1^\lambda)$; adversary $\mathcal{A}$ receives $k_{\mathrm{pub}}$.
    \item \textbf{Phase 1}: $\mathcal{A}$ submits chosen ciphertexts $c_i$ to the decapsulation oracle $O_{\mathrm{Dec}}(k_{\mathrm{pr}}, \cdot)$, receiving $k_i$ or $\perp$.
    \item \textbf{Challenge}: Challenger runs $(c^*, k_0) \leftarrow \mathrm{Enc}(k_{\mathrm{pub}})$; picks $b \leftarrow \{0,1\}$; if $b=1$, replaces $k_0$ with $k_1 \leftarrow \{0,1\}^{|k_0|}$. $\mathcal{A}$ receives $(c^*, k_b)$.
    \item \textbf{Phase 2}: $\mathcal{A}$ submits ciphertexts $c_j \neq c^*$ to $O_{\mathrm{Dec}}$.
    \item \textbf{Output}: $\mathcal{A}$ outputs $b' \in \{0,1\}$; wins if $b' = b$.
\end{enumerate}
The advantage is $\mathrm{Adv}^{\mathrm{IND\text{-}CCA2}}_{\Pi,\mathcal{A}}(\lambda) = |\Pr[b'=b] - 1/2|$. $\Pi$ is IND-CCA2 secure if this is $\mathrm{negl}(\lambda)$ for all PPT $\mathcal{A}$.
\end{definition}
This definition directly underlies the confidentiality guarantees for $k_{\mathrm{sh}}$ in both AmphiKey modes; Section~\ref{sec:security} reduces AKEM confidentiality to IND-CCA2 of ML-KEM-768 and X25519-DHKEM via hybrid games.
\section{Related Work and State-of-the-Art Analysis}
\label{sec:related}

\subsection{Hybrid Post-Quantum Key Exchange}
The transition to post-quantum cryptography has been supported by a wide range of hybrid KEM and key exchange designs. In this section, we show the most relevant work and identify the specific gaps AmphiKey addresses.

Recent hybrid post-quantum KEMs provide strong confidentiality but lack AmphiKey’s integrated authentication and smart-grid features. X-Wing~\cite{xwing_kem} cleanly combines X25519 and ML-KEM-768 with tight ROM security, 1,120-byte ciphertexts, and sub-millisecond performance, yet offers no built-in authentication, deniability, or SCA resistance—masking ML-KEM incurs $  >200\times  $ overhead on constrained devices. Hybrid HPKE~\cite{rfc9180} supports authenticated modes via sender key binding, but its current DHKEM is not post-quantum secure, emerging PQ variants lack deniability and SCA analysis, and formal work~\cite{alwen2021hpke} omits deniability. Signal’s PQXDH~\cite{signal_pqxdh} adds ML-KEM to X3DH for PQ forward secrecy while retaining informal deniability, but provides no formal PQ deniability proofs or SCA considerations. Apple’s PQ3~\cite{apple_pq3} integrates ML-KEM-768 into iMessage with classical ECDSA authentication, lacking PQ authentication, formal deniability, and IoT optimization.
The closest designs are Shadowfax and KEMTLS. Shadowfax~\cite{cryptoeprint:2025/154} builds a PQ deniability-preserving AKEM via NIKE, PQ KEM, and PQ ring signature with formal Real-or-Random proofs, but its 1,781-byte ciphertext exceeds AmphiKey’s Deniable Mode (1,152 bytes) and its ring signature has uncertain, likely difficult SCA posture. AmphiKey's HMAC-based deniable key exchange, derived from ephemeral secrets, is simpler, PRF-secure, and easier to mask. KEMTLS~\cite{kemtls2021} eliminates TLS 1.3 online signatures via KEM authentication for lower cost and forward secrecy, but offers no deniability (certified KEM keys identify parties) and no SCA analysis. AmphiKey’s Deniable Mode similarly avoids online signatures for efficiency while proving formal sender deniability; its Authenticated Mode uses online Raccoon signatures for non-repudiation which is essential for auditable smart grid traffic under regulatory needs that KEMTLS cannot meet.

\textbf{Hybrid combiners with formal proofs:} Bindel et al.~\cite{bindel2019hybrid} provide the foundational security analysis for hybrid KEM combiners, proving that combining IND-CCA2-secure KEMs via an appropriate combiner preserves IND-CCA2 security as long as at least one component is secure. The X-Wing~\cite{xwing_kem} combiner provides a tight security proof for the specific ML-KEM-768 + X25519 combination. AmphiKey's KEM combiner follows the same structure and inherits the same OR-security property; our proofs formalize this in the AmphiKey AKEM context.
\vspace{-3mm}
\subsection{Smart Grid Key Exchange Protocols}
Several domain-specific protocols address smart grid security. Mahmood et al.~\cite{chaudhry2017lightweight} proposed an ECC-based lightweight authentication scheme with low overhead, but it lacks quantum resistance. Bera and Sikdar~\cite{basuSmartgrid} introduced a post-quantum framework using CRYSTALS-Kyber and Dilithium, showing feasibility without SCA analysis or deniable authentication. Ahmad et al.~\cite{jian} (2025) combined blockchain, federated learning, and quantum-safe hybrid encryption for microgrids with strong threat modeling, yet omitted physical SCA countermeasures and deniability. Abdullah et al.~\cite{abdullah2025quantum} focused on NIST-standardized quantum-resistant hybrid encryption for smart grid IoT, but similarly lacks explicit SCA scope analysis and formal deniability proofs.
\subsection{SCA-Resistant Lattice Cryptography}
Masking lattice-based schemes remains challenging. Barthe et al.~\cite{Barthemasking} showed $  >200\times  $ overhead for masked GLP signatures (Dilithium predecessor), underscoring post-hoc masking's impracticality. Coron et al.~\cite{Coron2023Gadgets} reduced first-order Dilithium masking to $40$--$60\times$ overhead via specialized gadgets, still prohibitive for constrained devices. Raccoon DSA~\cite{delpino2024raccoon} addresses this by design: it eliminates rejection sampling and uses a masking-friendly structure for efficient high-order masking with $  O(d \log d)  $ overhead per share. Recent surveys on ML-KEM SCA~\cite{renitaSCA} highlight persistent KEM vulnerabilities, justifying AmphiKey's explicit scope analysis.
Table~\ref{tab:related_work_comparison} compares AmphiKey against key protocols across security and performance dimensions.

\begin{table*}[htbp]
\centering
\footnotesize
\caption{State-of-the-Art Comparison: Security Properties and Performance Characteristics}
\label{tab:related_work_comparison}
\renewcommand{\arraystretch}{1.1}
\begin{tabular}{@{}l c c c c c c c c@{}}
\toprule
\textbf{Protocol} & \textbf{PQ Conf.} & \textbf{Hybrid} & \textbf{Formal} & \textbf{Non-Rep.} & \textbf{SCA} & \textbf{Mode} & \textbf{Smart} & \textbf{Payload} \\
 & & \textbf{KEM} & \textbf{Deniability} & \textbf{Auth.} & \textbf{Scope} & \textbf{Downgrade} & \textbf{Grid} & \textbf{(Bytes)} \\
\midrule
X-Wing~\cite{xwing_kem}               & \checkmark & \checkmark & \texttimes & \texttimes & \texttimes & \texttimes & \texttimes & 1,120 \\
Hybrid HPKE~\cite{rfc9180}             & Partial    & Partial    & \texttimes & \texttimes & \texttimes & \texttimes & \texttimes & 1,152 \\
Signal PQXDH~\cite{signal_pqxdh}      & \checkmark & \checkmark & Informal   & \texttimes & \texttimes & N/A        & \texttimes & N/A \\
Apple PQ3~\cite{apple_pq3}            & \checkmark & \checkmark & \texttimes & Classical  & \texttimes & \texttimes & \texttimes & N/A \\
Shadowfax~\cite{cryptoeprint:2025/154}& \checkmark & \checkmark & \checkmark & \texttimes & \texttimes & \texttimes & \texttimes & 1,781 \\
Bera \& Sikdar~\cite{basuSmartgrid}   & \checkmark & \texttimes & \texttimes & \checkmark & \texttimes & \texttimes & \checkmark & N/A \\
Ahmad et al.~\cite{AHMAD2025126118}   & \checkmark & \checkmark & \texttimes & \texttimes & \texttimes & \texttimes & \checkmark & N/A \\
\midrule
\textbf{AmphiKey Auth}      & \checkmark & \checkmark & \texttimes  & \checkmark & \checkmark & \checkmark & \checkmark & 12,644 \\
\textbf{AmphiKey Deniable}  & \checkmark & \checkmark & \checkmark  & \texttimes & Partial    & \checkmark & \checkmark & 1,152 \\
\bottomrule
\end{tabular}
\captionsetup{font=small}
\caption*{\textbf{Legend.} PQ Conf.: Post-Quantum Confidentiality; Hybrid KEM: Hybrid PQ/classical KEM combiner; Formal Deniability: third-party sender deniability proven via simulation argument (distinct from receiver-side entity authentication); Non-Rep. Auth.: non-repudiable (EUF-CMA) authentication; SCA Scope: explicit SCA scope analysis with masking-friendly signing; Mode Downgrade: cryptographic protection against downgrade attacks; Smart Grid: evaluation on grid-representative hardware. Partial for Hybrid HPKE indicates PQ KEM not yet standardized in HPKE.}
\vspace{-5mm}
\end{table*}
\vspace{-3mm}
\section{System Architecture: AmphiKey Protocol}
\label{sec:methodology}
\subsection{System Architecture and Mode Selection}
\label{subsec:mode_selection}
AmphiKey targets a smart grid model with central servers (SCADA systems and utility control centers) and distributed IoT clients (smart meters, EV chargers, substation controllers, and PMUs). One party acts as the initiator (server in the typical smart grid model) and the other as the responder (client/device); roles are symmetric and interchangeable depending on the direction of communication. Long-term Raccoon DSA public keys are pre-shared and bound to device identities out-of-band (e.g., during device manufacturing or registration).
Mode selection occurs only at session initiation and remains fixed throughout the session. A single-bit \texttt{MODE} flag (\texttt{0} = Deniable, \texttt{1} = Authenticated) is included in the \emph{Server Hello} message and is \textbf{cryptographically bound} into all subsequent KDF inputs: specifically, it is concatenated into the HKDF context string before deriving $k_{\mathrm{sh}}$. This binding prevents an active adversary from flipping the mode bit undetected: any tampered \texttt{MODE} value causes key derivation to diverge from the intended session key, causing AEAD-authenticated application data to fail verification and the session to abort. The server proposes a mode based on a static policy table provisioned during device deployment. Safety-critical, auditable traffic (SCADA relay commands, firmware updates, substation control) uses Authenticated Mode, whereas high-volume, privacy-sensitive traffic (metering telemetry, EV charging session data, PMU readings) uses Deniable Mode. The client accepts or terminates; mid-session mode switching is not permitted, as it would introduce ambiguity in transcript integrity and potential replay vulnerabilities.
\subsection{AKEM Authentication Mode Construction}
\label{subsec:auth_mode}
\subsubsection*{Key Generation}
\begin{align*}
\text{Sender: }
&(k_{\mathrm{pr}_{s1}}, k_{\mathrm{pub}_{s1}})
    \leftarrow \mathrm{MLKEM.Gen}(1^\lambda),\\
& (k_{\mathrm{pr}_{s2}}, k_{\mathrm{pub}_{s2}})
    \leftarrow \mathrm{DHKEM.Gen}(1^\lambda),\\[4pt]
\text{Receiver: }
&(k_{\mathrm{pr}_{r1}}, k_{\mathrm{pub}_{r1}})
    \leftarrow \mathrm{MLKEM.Gen}(1^\lambda),\\
& (k_{\mathrm{pr}_{r2}}, k_{\mathrm{pub}_{r2}})
    \leftarrow \mathrm{DHKEM.Gen}(1^\lambda).
\end{align*}
Long-term Raccoon keypairs ($k_{\mathrm{pr}_{s,\mathrm{rac}}}, k_{\mathrm{pub}_{s,\mathrm{rac}}}$ for sender; similar for receiver) are pre-shared. The ML-KEM-768 and X25519-DHKEM keypairs are generated ephemerally per session.
\subsubsection*{Server Hello (Initiator)}
Let
\[
\mathit{SH}_s =
\big(
k_{\mathrm{pub}_{s1}} \,\|\,
k_{\mathrm{pub}_{s2}} \,\|\,
\texttt{MODE}
\big).
\]
The server computes:
\[
\mathrm{sig}_s \leftarrow
\mathrm{Rac.Sign}\big(
k_{\mathrm{pr}_{s,\mathrm{rac}}},\,
\mathit{SH}_s
\big),
\]
and transmits
\[
(\mathit{SH}_s,\, \mathrm{sig}_s).
\]
\subsubsection*{Encapsulation}
\emph{(Client, after verifying $\mathrm{sig}_s$)}
\begin{align*}
&\text{Verify }
\mathrm{Rac.Verify}\big(
    k_{\mathrm{pub}_{s,\mathrm{rac}}},\,
    \mathit{SH}_s,\,
    \mathrm{sig}_s
\big) = \top;\\
&\text{Abort if false.}\\[4pt]
&(c_1, k_1)
    \leftarrow \mathrm{MLKEM.Enc}(k_{\mathrm{pub}_{s1}}),\\
&(c_2, k_2)
    \leftarrow \mathrm{DHKEM.Enc}\big(
        k_{\mathrm{pr}_{c2}},\,
        k_{\mathrm{pub}_{s2}}
    \big),\\
&c \leftarrow c_1 \,\|\, c_2,\\[4pt]
&\mathrm{sig}_c \leftarrow
\mathrm{Rac.Sign}\big(
    k_{\mathrm{pr}_{c,\mathrm{rac}}},\,
    c \,\|\, \mathit{SH}_s
\big),\\[6pt]
&k_{\mathrm{sh}} \leftarrow
\mathrm{HKDF\text{-}SHA256}\big(
    k_1 \,\|\, k_2 \,\|\, c \,\|\, \mathrm{sig}_s \\
&\quad \|\, \mathrm{sig}_c
    \,\|\, k_{\mathrm{pub}_{s,\mathrm{rac}}}
    \,\|\, k_{\mathrm{pub}_{c,\mathrm{rac}}}
    \,\|\, \texttt{MODE}
\big),\\[4pt]
&\text{return }
(c,\, \mathrm{sig}_c,\, k_{\mathrm{sh}}).
\end{align*}
\subsubsection*{Decapsulation}
\emph{(Server)}
\begin{align*}
&\text{Verify }
\mathrm{Rac.Verify}\big(
    k_{\mathrm{pub}_{c,\mathrm{rac}}},\,
    c \,\|\, \mathit{SH}_s,\,
    \mathrm{sig}_c
\big) = \top;\\
&\text{Abort if false.}\\[4pt]
&k_1 \leftarrow
    \mathrm{MLKEM.Dec}(k_{\mathrm{pr}_{s1}}, c_1),\\
&k_2 \leftarrow
    \mathrm{DHKEM.Dec}(k_{\mathrm{pr}_{s2}}, c_2),\\
&\text{If } k_1 = \perp
   \text{ or } k_2 = \perp:
   \text{ return } \perp,\\[6pt]
&k_{\mathrm{sh}} \leftarrow
\mathrm{HKDF\text{-}SHA256}\big(
    k_1 \,\|\, k_2 \,\|\, c \,\|\, \mathrm{sig}_s \\
&\quad \|\, \mathrm{sig}_c
    \,\|\, k_{\mathrm{pub}_{s,\mathrm{rac}}}
    \,\|\, k_{\mathrm{pub}_{c,\mathrm{rac}}}
    \,\|\, \texttt{MODE}
\big).
\end{align*}
\textbf{Design Rationale:} Two Raccoon signatures bind the full handshake transcript in both directions: $\mathrm{sig}_s$ authenticates the server's ephemeral KEM public keys and mode proposal; $\mathrm{sig}_c$ authenticates the client's ciphertexts to the server's Hello. Including the full Server Hello in the client's signed message prevents unknown-key-share attacks and authenticates the server's ephemeral material to the client before encapsulation proceeds. The `OR' confidentiality structure ensures that $k_{\mathrm{sh}}$ remains hidden as long as at least one KEM remains unbroken. The `AND' authenticity requires both signatures to be verified and both KEMs to be successfully decapsulated. The \texttt{MODE} flag and both pre-shared public keys are included in the HKDF transcript, preventing both mode downgrade and identity substitution attacks. The complete handshake flow is illustrated in Figure~\ref{fig:handshake_auth_mode}.
\vspace{-5mm}
\subsection{AKEM Deniable Mode Construction}
\label{subsec:den_mode}
\subsubsection*{Key Generation}
Identical to the Authentication Mode except no Raccoon keypair is generated. All KEM keypairs are ephemeral per session.
\subsubsection*{Server Hello}
To ensure replay resistance without relying solely on ciphertext freshness, the server generates a fresh random challenge nonce $r_s \leftarrow \{0,1\}^{128}$ and transmits it alongside its ephemeral public keys in the Server Hello:
\[
\mathit{SH}_s^{\mathrm{den}} = (k_{\mathrm{pub}_{r1}},\, k_{\mathrm{pub}_{r2}},\, r_s,\, \texttt{MODE}).
\]
The server retains $r_s$ and discards it after the session; it must not be reused across sessions.
\subsubsection*{Encapsulation} $(k_{\mathrm{pr}_s}, k_{\mathrm{pub}_r}, r_s)$:
\begin{align*}
&(c_1, k_1) \leftarrow \mathrm{MLKEM.Enc}(k_{\mathrm{pub}_{r1}}),\\
&(c_2, k_2) \leftarrow \mathrm{DHKEM.Enc}(k_{\mathrm{pr}_{s2}}, k_{\mathrm{pub}_{r2}}),\\
&c \leftarrow c_1 \| c_2,\\
&k_{\mathrm{auth}} \leftarrow \mathrm{HKDF\text{-}SHA256}(k_1 \| k_2 \| r_s,\, \text{``auth''}),\\
&\mathrm{tag} \leftarrow \mathrm{HMAC\text{-}SHA256}(k_{\mathrm{auth}},\, c \| r_s \| \texttt{MODE}),\\
&k_{\mathrm{sh}} \leftarrow \mathrm{HKDF\text{-}SHA256}(k_1 \| k_2 \| c \| r_s \| k_{\mathrm{pub}_r} \| \texttt{MODE}),\\
&\text{return } (c,\, \mathrm{tag},\, k_{\mathrm{sh}}).
\end{align*}
\subsubsection*{Decapsulation} $(k_{\mathrm{pr}_r}, r_s, \mathrm{tag}, c)$:
\begin{align*}
k_1 &\leftarrow \mathrm{MLKEM.Dec}(k_{\mathrm{pr}_{r1}},\, c_1), \\
k_2 &\leftarrow \mathrm{DHKEM.Dec}(k_{\mathrm{pr}_{r2}},\, c_2), \\[4pt]
\text{If } &k_1 = \perp \;\text{or}\; k_2 = \perp \\
&\text{: return } \perp, \\[6pt]
k_{\mathrm{auth}} 
&\leftarrow \mathrm{HKDF\text{-}SHA256}\!\left(
    k_1 \,\|\, k_2 \,\|\, r_s,\,
    \text{``auth''}
\right), \\[6pt]
\text{If } 
&\mathrm{HMAC\text{-}SHA256}\!\left(
    k_{\mathrm{auth}},\,
    c \,\|\, r_s \,\|\, \texttt{MODE}
\right)
\neq \mathrm{tag} \\
&\text{: return } \perp, \\[6pt]
k_{\mathrm{sh}} 
&\leftarrow \mathrm{HKDF\text{-}SHA256}\!\left(
    k_1 \,\|\, k_2 \,\|\, c \,\|\, r_s \,\|\, 
    k_{\mathrm{pub}_r} \,\|\, \texttt{MODE}
\right).
\end{align*}

\begin{figure*}[!t]
\centering
\tikzset{
  sbox/.style={rectangle,rounded corners=4pt,text width=8.5cm,font=\normalsize,
    inner sep=6pt,align=left,line width=1.8pt,fill=stepbg!80!white,draw=boxborder!60},
  cbox/.style={rectangle,rounded corners=4pt,text width=8.5cm,font=\normalsize,
    inner sep=6pt,align=left,line width=2.2pt,fill=critbg,draw=orange!80!black},
  amphdr/.style={rectangle,rounded corners=5pt,minimum width=8.8cm,minimum height=1.0cm,
    text centered,font=\normalsize\bfseries\sffamily,text=white,inner sep=5pt},
  amparr/.style={-{Stealth[length=10pt,width=7pt]},line width=2.0pt,color=arrowcol},
  amparrL/.style={{Stealth[length=10pt,width=7pt]}-,line width=2.0pt,color=arrowcol},
  amptd/.style={densely dashed,line width=0.8pt,color=gray!45},
}
\resizebox{0.7\textwidth}{!}{%
\begin{tikzpicture}
\def\CX{2.5}\def\SX{13.5}\def\MX{8.0}
\draw[amptd](\CX,-0.45)--(\CX,-24.0);
\draw[amptd](\SX,-0.45)--(\SX,-24.0);

\node[amphdr,fill=headerbg] at(\CX,0)
  {Client \normalfont\large(Smart Meter)};
\node[amphdr,fill=headerbg!75!black] at(\SX,0)
  {Server \normalfont\large(SCADA)};

\node[sbox,anchor=north] at(\CX,-0.45){%
  \textbf{Key Generation}\\[1pt]
  $(k_{\mathrm{pr}_{r1}},k_{\mathrm{pub}_{r1}})\leftarrow\mathrm{MLKEM.Gen}$\\[0.5pt]
  $(k_{\mathrm{pr}_{r2}},k_{\mathrm{pub}_{r2}})\leftarrow\mathrm{DHKEM.Gen}$\\[0.5pt]
  \textit{Long-term:} $k_{\mathrm{pr}_{c,\mathrm{rac}}},k_{\mathrm{pub}_{c,\mathrm{rac}}}$};

\node[sbox,anchor=north] at(\SX,-0.45){%
  \textbf{Key Generation}\\[1pt]
  $(k_{\mathrm{pr}_{s1}},k_{\mathrm{pub}_{s1}})\leftarrow\mathrm{MLKEM.Gen}$\\[0.5pt]
  $(k_{\mathrm{pr}_{s2}},k_{\mathrm{pub}_{s2}})\leftarrow\mathrm{DHKEM.Gen}$\\[0.5pt]
  \textit{Long-term:} $k_{\mathrm{pr}_{s,\mathrm{rac}}},k_{\mathrm{pub}_{s,\mathrm{rac}}}$\\[2pt]
  $\mathit{SH}_s\leftarrow k_{\mathrm{pub}_{s1}}\|k_{\mathrm{pub}_{s2}}\|\texttt{MODE}$\\[0.5pt]
  $\mathrm{sig}_s\leftarrow\mathrm{Rac.Sign}(k_{\mathrm{pr}_{s,\mathrm{rac}}},\mathit{SH}_s)$};

\def\MA{-4.25}
\draw[amparrL](\CX+0.15,\MA)--(\SX-0.15,\MA);
\node[fill=white,inner sep=2pt,font=\normalsize\bfseries\sffamily,text=arrowcol]
  at(\MX,\MA+0.35){Server Hello};
\node[fill=white,inner sep=1.5pt,font=\normalsize,text=arrowcol]
  at(\MX,\MA-0.40)
  {$k_{\mathrm{pub}_{s1}},\;k_{\mathrm{pub}_{s2}},\;\mathrm{sig}_s,\;k_{\mathrm{pub}_{s,\mathrm{rac}}},\;\texttt{MODE}{=}1$};

\node[cbox,anchor=north] at(\CX,-5.0){%
  \textbf{Verify}\enspace
    $\mathrm{Rac.Verify}(k_{\mathrm{pub}_{s,\mathrm{rac}}},\mathit{SH}_s,\mathrm{sig}_s)=\top$\\[0.3pt]
  \textit{Abort if false.}\\[2pt]
  \textbf{Encapsulate}\\[0.5pt]
  $(c_1,k_1)\leftarrow\mathrm{MLKEM.Enc}(k_{\mathrm{pub}_{s1}})$\\[0.5pt]
  $(c_2,k_2)\leftarrow\mathrm{DHKEM.Enc}(k_{\mathrm{pr}_{c2}},k_{\mathrm{pub}_{s2}})$\\[0.5pt]
  $c\leftarrow c_1\|c_2$\\[2pt]
  \textbf{Sign}\enspace{\normalsize\textit{(Raccoon, SCA-masked)}}\\[0.5pt]
  $\mathrm{sig}_c\leftarrow\mathrm{Rac.Sign}(k_{\mathrm{pr}_{c,\mathrm{rac}}},c\|\mathit{SH}_s)$\\[2pt]
  \textbf{Derive}\\[0.5pt]
  $k_{\mathrm{sh}}\leftarrow\mathrm{HKDF}(k_1\|k_2\|c\|\mathrm{sig}_s$\\[0.3pt]
  $\quad\|\mathrm{sig}_c\|k_{\mathrm{pub}_{s,\mathrm{rac}}}\|k_{\mathrm{pub}_{c,\mathrm{rac}}}\|\texttt{M})$};

\def\MB{-11.20}
\draw[amparr](\CX+0.15,\MB)--(\SX-0.15,\MB);
\node[fill=white,inner sep=2pt,font=\normalsize\bfseries\sffamily,text=arrowcol]
  at(\MX,\MB+0.35){Encapsulation};
\node[fill=white,inner sep=1.5pt,font=\normalsize,text=arrowcol]
  at(\MX,\MB-0.35){$c,\;\mathrm{sig}_c,\;k_{\mathrm{pub}_{c,\mathrm{rac}}}$};

\node[cbox,anchor=north] at(\SX,-12.0){%
  \textbf{Verify}\enspace
    $\mathrm{Rac.Verify}(k_{\mathrm{pub}_{c,\mathrm{rac}}},c\|\mathit{SH}_s,\mathrm{sig}_c)=\top$\\[0.3pt]
  \textit{Abort if false.}\\[2pt]
  \textbf{Decapsulate}\\[0.5pt]
  $k_1\leftarrow\mathrm{MLKEM.Dec}(k_{\mathrm{pr}_{s1}},c_1)$\\[0.5pt]
  $k_2\leftarrow\mathrm{DHKEM.Dec}(k_{\mathrm{pr}_{s2}},c_2)$\\[0.5pt]
  \textit{If $k_1{=}\perp$ or $k_2{=}\perp$: abort.}\\[2pt]
  \textbf{Derive}\\[0.5pt]
  $k_{\mathrm{sh}}\leftarrow\mathrm{HKDF}(k_1\|k_2\|c\|\mathrm{sig}_s$\\[0.3pt]
  $\quad\|\mathrm{sig}_c\|k_{\mathrm{pub}_{s,\mathrm{rac}}}\|k_{\mathrm{pub}_{c,\mathrm{rac}}}\|\texttt{M})$};

\draw[dotted,line width=1.0pt,gray!55]
  (4.0,-16.5)--(8.5,-16.5)
  node[midway,fill=white,font=\normalsize\itshape,text=gray!55,inner sep=1pt]
  {same $k_{\mathrm{sh}}$};

\node[sbox,anchor=north] at(\SX,-17.0){%
  \textbf{Encrypt}\\[0.5pt]
  $C_R\leftarrow\mathrm{Ascon.Enc}(k_{\mathrm{sh}},n_R,P_R)$};

\def\MC{-19.0}
\draw[amparrL](\CX+0.15,\MC)--(\SX-0.15,\MC);
\node[fill=white,inner sep=1.5pt,font=\normalsize,text=arrowcol]
  at(\MX,\MC+0.32){$\xleftarrow{\;\text{Enc.\ Request:}\;n_R,C_R\;}$};

\node[sbox,anchor=north] at(\CX,-19.5){%
  $P_R\leftarrow\mathrm{Ascon.Dec}(k_{\mathrm{sh}},n_R,C_R)$\\[0.5pt]
  $C_S\leftarrow\mathrm{Ascon.Enc}(k_{\mathrm{sh}},n_S,P_S)$};

\def\MD{-21.5}
\draw[amparr](\CX+0.15,\MD)--(\SX-0.15,\MD);
\node[fill=white,inner sep=1.5pt,font=\normalsize,text=arrowcol]
  at(\MX,\MD+0.32){$\xrightarrow{\;\text{Enc.\ Response:}\;n_S,C_S\;}$};

\node[sbox,anchor=north] at(\SX,-22.0){%
  $P_S\leftarrow\mathrm{Ascon.Dec}(k_{\mathrm{sh}},n_S,C_S)$};

\def\PY{-24.00}
\node[rectangle,rounded corners=4pt,fill=clientbg,draw=boxborder,line width=1.2pt,
  text width=3.6cm,font=\normalsize,inner sep=6pt,align=center] at(0.80,\PY){%
  \textcolor{boxborder}{\textbf{OR-Conf.}}\\[2pt]
  $k_{\mathrm{sh}}$ secret if\\either KEM holds};
\node[rectangle,rounded corners=4pt,fill=stepbg,draw=gray!70,line width=1.2pt,
  text width=3.8cm,font=\normalsize,inner sep=6pt,align=center] at(\MX,\PY){%
  \textcolor{gray!60!black}{\textbf{Downgrade Prot.}}\\[2pt]
  \texttt{MODE}$=$1 in HKDF;\\tamper $\Rightarrow$ abort};
\node[rectangle,rounded corners=4pt,fill=serverbg,draw=headerbg!80!black,line width=1.2pt,
  text width=3.6cm,font=\normalsize,inner sep=6pt,align=center] at(15.20,\PY){%
  \textcolor{headerbg!80!black}{\textbf{AND-Auth.}}\\[2pt]
  Both sigs valid\\$+$ both KEMs ok};

\node[font=\normalsize\bfseries\sffamily,text=headerbg] at(\MX,-25.60){\textbf{Authenticated Mode}};
\end{tikzpicture}%
}
\caption{AmphiKey handshake flow in \textbf{Authenticated Mode} ({\small\texttt{M}}$=${\small\texttt{MODE}}).}
\label{fig:handshake_auth_mode}
\vspace{-3mm}
\end{figure*}

\begin{figure*}[!t]
\centering
\tikzset{
  sbox/.style={rectangle,rounded corners=4pt,text width=8.5cm,font=\normalsize,
    inner sep=6pt,align=left,line width=1.8pt,fill=stepbg!80!white,draw=boxborder!60},
  cbox/.style={rectangle,rounded corners=4pt,text width=8.5cm,font=\normalsize,
    inner sep=6pt,align=left,line width=2.2pt,fill=critbg,draw=orange!80!black},
  amphdr/.style={rectangle,rounded corners=5pt,minimum width=8.8cm,minimum height=1.0cm,
    text centered,font=\normalsize\bfseries\sffamily,text=white,inner sep=5pt},
  amparr/.style={-{Stealth[length=10pt,width=7pt]},line width=2.0pt,color=arrowcol},
  amparrL/.style={{Stealth[length=10pt,width=7pt]}-,line width=2.0pt,color=arrowcol},
  amptd/.style={densely dashed,line width=0.8pt,color=gray!45},
}
\resizebox{0.7\textwidth}{!}{%
\begin{tikzpicture}
\def\CX{2.5}\def\SX{13.5}\def\MX{8.0}
\draw[amptd](\CX,-0.45)--(\CX,-24.0);
\draw[amptd](\SX,-0.45)--(\SX,-24.0);

\node[amphdr,fill=headerbg] at(\CX,0)
  {Client \normalfont\large(Smart Meter)};
\node[amphdr,fill=headerbg!75!black] at(\SX,0)
  {Server \normalfont\large(SCADA)};

\node[sbox,anchor=north] at(\CX,-0.45){%
  \textbf{Key Generation}\enspace{\normalsize\textit{(ephemeral)}}\\[1pt]
  $(k_{\mathrm{pr}_{r1}},k_{\mathrm{pub}_{r1}})\leftarrow\mathrm{MLKEM.Gen}$\\[0.5pt]
  $(k_{\mathrm{pr}_{r2}},k_{\mathrm{pub}_{r2}})\leftarrow\mathrm{DHKEM.Gen}$\\[0.5pt]
  \textit{No long-term signing key}};

\node[sbox,anchor=north] at(\SX,-0.45){%
  \textbf{Key Generation}\enspace{\normalsize\textit{(ephemeral)}}\\[1pt]
  $(k_{\mathrm{pr}_{s2}},k_{\mathrm{pub}_{s2}})\leftarrow\mathrm{DHKEM.Gen}$\\[0.5pt]
  $r_s\xleftarrow{\$}\{0,1\}^{128}$\enspace{\normalsize(session nonce)}\\[0.5pt]
  \textit{No long-term signing key}};

\def\MA{-3.80}
\draw[amparrL](\CX+0.15,\MA)--(\SX-0.15,\MA);
\node[fill=white,inner sep=2pt,font=\normalsize\bfseries\sffamily,text=arrowcol]
  at(\MX,\MA+0.35){Server Hello};
\node[fill=white,inner sep=1.5pt,font=\normalsize,text=arrowcol]
  at(\MX,\MA-0.40)
  {$\mathit{SH}_s^{\mathrm{den}}=(k_{\mathrm{pub}_{r1}},k_{\mathrm{pub}_{r2}},r_s,\texttt{MODE}{=}0)$};

\node[cbox,anchor=north] at(\CX,-4.50){%
  \textbf{Encapsulate}\\[0.5pt]
  $(c_1,k_1)\leftarrow\mathrm{MLKEM.Enc}(k_{\mathrm{pub}_{r1}})$\\[0.5pt]
  $(c_2,k_2)\leftarrow\mathrm{DHKEM.Enc}(k_{\mathrm{pr}_{s2}},k_{\mathrm{pub}_{r2}})$\\[0.5pt]
  $c\leftarrow c_1\|c_2$\\[2pt]
  \textbf{Auth Key}\\[0.5pt]
  $k_{\mathrm{auth}}\leftarrow\mathrm{HKDF}(k_1\|k_2\|r_s,\text{``auth''})$\\[2pt]
  \textbf{HMAC Tag}\enspace{\normalsize\textit{(replaces signature)}}\\[0.5pt]
  $\mathrm{tag}\leftarrow\mathrm{HMAC}(k_{\mathrm{auth}},c\|r_s\|\texttt{MODE})$\\[2pt]
  \textbf{Derive}\\[0.5pt]
  $k_{\mathrm{sh}}\leftarrow\mathrm{HKDF}(k_1\|k_2\|c\|r_s\|k_{\mathrm{pub}_r}\|\texttt{M})$};

\def\MB{-10.60}
\draw[amparr](\CX+0.15,\MB)--(\SX-0.15,\MB);
\node[fill=white,inner sep=2pt,font=\normalsize\bfseries\sffamily,text=arrowcol]
  at(\MX,\MB+0.35){Encapsulation};
\node[fill=white,inner sep=1.5pt,font=\normalsize,text=arrowcol]
  at(\MX,\MB-0.35){$c=(c_1\|c_2),\;\mathrm{tag}$};

\node[cbox,anchor=north] at(\SX,-11.40){%
  \textbf{Decapsulate}\\[0.5pt]
  $k_1\leftarrow\mathrm{MLKEM.Dec}(k_{\mathrm{pr}_{r1}},c_1)$\\[0.5pt]
  $k_2\leftarrow\mathrm{DHKEM.Dec}(k_{\mathrm{pr}_{r2}},c_2)$\\[0.5pt]
  \textit{If $k_1{=}\perp$ or $k_2{=}\perp$: abort.}\\[2pt]
  \textbf{Verify HMAC Tag}\\[0.5pt]
  $k_{\mathrm{auth}}\leftarrow\mathrm{HKDF}(k_1\|k_2\|r_s,\text{``auth''})$\\[0.5pt]
  \textit{If}\enspace$\mathrm{HMAC}(k_{\mathrm{auth}},c\|r_s\|\texttt{M})\neq\mathrm{tag}$: abort.\\[2pt]
  \textbf{Derive}\\[0.5pt]
  $k_{\mathrm{sh}}\leftarrow\mathrm{HKDF}(k_1\|k_2\|c\|r_s\|k_{\mathrm{pub}_r}\|\texttt{M})$};

\draw[dotted,line width=1.0pt,gray!55]
  (4.0,-15.90)--(8.5,-15.90)
  node[midway,fill=white,font=\normalsize\itshape,text=gray!55,inner sep=1pt]
  {same $k_{\mathrm{sh}}$};

\node[sbox,anchor=north] at(\SX,-16.40){%
  \textbf{Encrypt}\\[0.5pt]
  $C_R\leftarrow\mathrm{Ascon.Enc}(k_{\mathrm{sh}},n_R,P_R)$};

\def\MC{-18.40}
\draw[amparrL](\CX+0.15,\MC)--(\SX-0.15,\MC);
\node[fill=white,inner sep=1.5pt,font=\normalsize,text=arrowcol]
  at(\MX,\MC+0.32){$\xleftarrow{\;\text{Enc.\ Request:}\;n_R,C_R\;}$};

\node[sbox,anchor=north] at(\CX,-18.90){%
  $P_R\leftarrow\mathrm{Ascon.Dec}(k_{\mathrm{sh}},n_R,C_R)$\\[0.5pt]
  $C_S\leftarrow\mathrm{Ascon.Enc}(k_{\mathrm{sh}},n_S,P_S)$};

\def\MD{-20.90}
\draw[amparr](\CX+0.15,\MD)--(\SX-0.15,\MD);
\node[fill=white,inner sep=1.5pt,font=\normalsize,text=arrowcol]
  at(\MX,\MD+0.32){$\xrightarrow{\;\text{Enc.\ Response:}\;n_S,C_S\;}$};

\node[sbox,anchor=north] at(\SX,-21.40){%
  $P_S\leftarrow\mathrm{Ascon.Dec}(k_{\mathrm{sh}},n_S,C_S)$};

\def\PY{-24.00}
\node[rectangle,rounded corners=4pt,fill=clientbg,draw=boxborder,line width=1.2pt,
  text width=3.6cm,font=\normalsize,inner sep=6pt,align=center] at(0.80,\PY){%
  \textcolor{boxborder}{\textbf{Sender Deniab.}}\\[2pt]
  $\mathrm{Sim}(k_{\mathrm{pr}_r},r_s)$:\\indisting.\ transcript};
\node[rectangle,rounded corners=4pt,fill=stepbg,draw=gray!70,line width=1.2pt,
  text width=3.8cm,font=\normalsize,inner sep=6pt,align=center] at(\MX,\PY){%
  \textcolor{gray!60!black}{\textbf{Replay Resistance}}\\[2pt]
  Fresh $r_s$ per session;\\replayed tag fails};
\node[rectangle,rounded corners=4pt,fill=serverbg,draw=headerbg!80!black,line width=1.2pt,
  text width=3.6cm,font=\normalsize,inner sep=6pt,align=center] at(15.20,\PY){%
  \textcolor{headerbg!80!black}{\textbf{93\% Less CPU}}\\[2pt]
  ${\sim}0.41$\,ms vs\\${\sim}4.8$\,ms Auth};

\node[font=\normalsize\bfseries\sffamily,text=headerbg!75!black] at(\MX,-25.60){\textbf{Deniable Mode}};
\end{tikzpicture}%
}
\caption{AmphiKey handshake flow in \textbf{Deniable Mode} ({\small\texttt{M}}$=${\small\texttt{MODE}}).}
\label{fig:handshake_deniable_mode}
\vspace{-3mm}
\end{figure*}

\textbf{Design Rationale and Replay Resistance:} The server-generated challenge nonce $r_s$ is the primary replay-resistance mechanism. Since $r_s$ is freshly sampled per session and included in both the HMAC tag and the $k_{\mathrm{sh}}$ derivation, a replayed transcript $(c, \mathrm{tag})$ from a prior session will fail HMAC verification under any new $r_s'$ the server generates. The complete Deniable Mode flow is illustrated in Figure~\ref{fig:handshake_deniable_mode}.
\vspace{-3mm}
\subsection{Scope and Limitations of SCA Protection}
\label{subsec:sca_scope}
Raccoon’s masking countermeasures protect \emph{only} the signature generation step ($  \mathrm{sig} \leftarrow \mathrm{Rac.Sign}(k_{\mathrm{pr}_{s,\mathrm{rac}}}, c)  $), safeguarding the long-term secret $  k_{\mathrm{pr}_{s,\mathrm{rac}}}  $ (and $  k_{\mathrm{pr}_{c,\mathrm{rac}}}  $) during signing. All other operations—ML-KEM-768 and X25519-DHKEM encapsulation/decapsulation, HKDF derivation, and HMAC computation—are \emph{unmasked} in our reference implementation. A physically present adversary can thus use first-order DPA or EM attacks to extract ephemeral secrets $  (k_1, k_2)  $ or session keys $  k_{\mathrm{sh}}  $.
Mitigating these residual vulnerabilities requires separate countermeasures: masked NTTs for ML-KEM, additional masking for X25519 (beyond libsodium’s constant-time design), and noise injection or hardware security modules across the handshake path. These fall outside protocol-layer scope and are deployment responsibilities.
Although $  \mathrm{sig}_s  $ and $  \mathrm{sig}_c  $ are included in HKDF inputs, they are public and provide no confidentiality for $  k_{\mathrm{sh}}  $ once $  (k_1, k_2)  $ are leaked using SCA where the attacker can recompute $  k_{\mathrm{sh}}  $ directly from the transcript. Thus, Raccoon’s SCA protection is strictly limited to preserving long-term signing-key security during signing operations. Without masking the KEM components themselves, ephemeral extraction enables full session-key recovery.

\section{Security Analysis}
\label{sec:security}
We present formal security proofs for both AmphiKey modes using a sequence-of-games methodology. All reductions are concrete, bounding the protocol advantage as a sum of advantages against underlying primitive security games. Let $\mathrm{Adv}^X_P$ denote the advantage of the best PPT adversary against property $X$ of primitive $P$.
\textbf{Hybrid combiner security.} The `OR' confidentiality of the two-KEM combiner $k_{\mathrm{sh}} = \mathrm{HKDF}(k_1 \| k_2 \| \cdots)$ is formally supported by the hybrid combiner analysis of Bindel et al.~\cite{bindel2019hybrid} and the $\mathrm{X\text{-}Wing}$ security analysis of Connolly et al.~\cite{xwing_kem}: when $H$ is a PRF and at least one of $(k_1, k_2)$ is computationally hidden, $k_{\mathrm{sh}}$ is indistinguishable from random. Our proofs instantiate this result.
\vspace{-3mm}
\subsection{Authenticated Mode: Formal IND-CCA2 Game}
\label{subsec:auth_game}
The ``ciphertext'' for the Auth Mode AKEM is $(c, \mathrm{sig}_s, \mathrm{sig}_c)$ (server Hello signature and client ciphertext signature) and the ``session key'' is $k_{\mathrm{sh}}$.
\textbf{AKEM-Auth-IND-CCA2 Game:}
\begin{enumerate}[leftmargin=*]
    \item \textbf{Setup}: Challenger generates server keys $(k_{\mathrm{pr}_{s1}}, k_{\mathrm{pub}_{s1}})$, $(k_{\mathrm{pr}_{s2}}, k_{\mathrm{pub}_{s2}})$, $(k_{\mathrm{pr}_{s,\mathrm{rac}}}, k_{\mathrm{pub}_{s,\mathrm{rac}}})$ and client keys $(k_{\mathrm{pr}_{c,\mathrm{rac}}}, k_{\mathrm{pub}_{c,\mathrm{rac}}})$. Adversary $\mathcal{A}$ receives all public keys.
    \item \textbf{Phase 1 (Decapsulation Oracle)}: $\mathcal{A}$ queries $O_{\mathrm{Dec}}(\mathit{SH}_s', \mathrm{sig}_s', c', \mathrm{sig}_c')$. Challenger: (i) verifies both $\mathrm{Rac.Verify}(k_{\mathrm{pub}_{s,\mathrm{rac}}}, \mathit{SH}_s', \mathrm{sig}_s')$ and $\mathrm{Rac.Verify}(k_{\mathrm{pub}_{c,\mathrm{rac}}}, c' \| \mathit{SH}_s', \mathrm{sig}_c')$, returning $\perp$ on failure; (ii) decapsulates $(k_1', k_2')$ from $(c_1', c_2')$; (iii) returns $k_{\mathrm{sh}}' = \mathrm{HKDF}(k_1' \| k_2' \| c' \| \mathrm{sig}_s' \| \mathrm{sig}_c' \| k_{\mathrm{pub}_{s,\mathrm{rac}}} \| k_{\mathrm{pub}_{c,\mathrm{rac}}} \| \texttt{MODE})$.
    \item \textbf{Challenge}: Challenger generates $\mathit{SH}_s^*$, $\mathrm{sig}_s^*$, $(c_1^*, k_1^*)$, $(c_2^*, k_2^*)$, $\mathrm{sig}_c^*$, $k_{\mathrm{sh}_0}$ per the protocol. If $b=0$: challenge key $= k_{\mathrm{sh}_0}$; if $b=1$: uniform random. $\mathcal{A}$ receives challenge transcript.
    \item \textbf{Phase 2}: $\mathcal{A}$ queries $O_{\mathrm{Dec}}$ with $({\mathit{SH}_s}', \mathrm{sig}_s', c', \mathrm{sig}_c') \neq (\mathit{SH}_s^*, \mathrm{sig}_s^*, c^*, \mathrm{sig}_c^*)$.
    \item \textbf{Guess}: $\mathcal{A}$ outputs $b' \in \{0,1\}$; wins if $b' = b$.
\end{enumerate}
Advantage: $\mathrm{Adv}^{\mathrm{IND\text{-}CCA2}}_{\mathrm{Auth},\mathcal{A}} = |\Pr[b' = b] - 1/2|$.
\vspace{-3mm}
\subsection{Authenticated Mode: Sequence-of-Games Proof}
\begin{theorem}[Auth Mode IND-CCA2]
\label{thm:auth_conf}
If ML-KEM-768 and X25519-DHKEM are IND-CCA2 secure, Raccoon DSA is EUF-CMA secure (for both server and client signing keys), and HKDF-SHA-256 is a secure PRF, then $k_{\mathrm{sh}}$ in the Authenticated Mode is IND-CCA2 secure. For any PPT adversary $\mathcal{A}$:
\[
\begin{aligned}
\mathrm{Adv}^{\mathrm{IND\text{-}CCA2}}_{\mathrm{Auth},\mathcal{A}}
\;\leq\;&
2 \cdot \mathrm{Adv}^{\mathrm{EUF\text{-}CMA}}_{\mathrm{Rac}} \\
&+ \mathrm{Adv}^{\mathrm{IND\text{-}CCA2}}_{\mathrm{MLKEM}} \\
&+ \mathrm{Adv}^{\mathrm{IND\text{-}CCA2}}_{\mathrm{DHKEM}} \\
&+ \mathrm{Adv}^{\mathrm{PRF}}_{H}.
\end{aligned}
\]
\end{theorem}
\begin{proof}
We define games $G_0, \ldots, G_4$ bounding the difference in adversarial winning probability between consecutive games.\\
\textbf{Game $G_0$ (Real):} $\mathrm{Adv}_\mathcal{A}[G_0] = \mathrm{Adv}^{\mathrm{IND\text{-}CCA2}}_{\mathrm{Auth},\mathcal{A}}$.\\
\textbf{Game $G_1$ (Server EUF-CMA Abort):} Abort if $\mathcal{A}$ queries $O_{\mathrm{Dec}}$ with $(\mathit{SH}_s^*, \mathrm{sig}_s', \cdot, \cdot)$ where $\mathrm{sig}_s' \neq \mathrm{sig}_s^*$ but $\mathrm{Rac.Verify}(k_{\mathrm{pub}_{s,\mathrm{rac}}}, \mathit{SH}_s^*, \mathrm{sig}_s') = \top$---an EUF-CMA forgery on the server's signing key.
\emph{Reduction $B_{\mathrm{EUF}_s}$}: Receives server's Raccoon public key and signing oracle; uses them to generate $\mathrm{sig}_s^*$ and detect forgeries on $\mathit{SH}_s^*$.
\[ |\Pr[\mathrm{Win}_{G_0}] - \Pr[\mathrm{Win}_{G_1}]| \leq \mathrm{Adv}^{\mathrm{EUF\text{-}CMA}}_{\mathrm{Rac}, B_{\mathrm{EUF}_s}}. \]
\textbf{Game $G_{1.5}$ (Client EUF-CMA Abort):} Additionally abort if $\mathcal{A}$ queries $O_{\mathrm{Dec}}$ with $(\cdot, \cdot, c^*, \mathrm{sig}_c')$ where $\mathrm{sig}_c' \neq \mathrm{sig}_c^*$ but $\mathrm{Rac.Verify}(k_{\mathrm{pub}_{c,\mathrm{rac}}}, c^* \| \mathit{SH}_s^*, \mathrm{sig}_c') = \top$---an EUF-CMA forgery on the client's signing key.
\emph{Reduction $B_{\mathrm{EUF}_c}$}: Analogously uses the client's Raccoon signing oracle.
\[ |\Pr[\mathrm{Win}_{G_1}] - \Pr[\mathrm{Win}_{G_{1.5}}]| \leq \mathrm{Adv}^{\mathrm{EUF\text{-}CMA}}_{\mathrm{Rac}, B_{\mathrm{EUF}_c}}. \]
\textbf{Game $G_2$ (ML-KEM Randomization):} Replace challenge key $k_1^*$ with $k_{1,R} \leftarrow \{0,1\}^{256}$.
\emph{Reduction $B_{\mathrm{KEM1}}$}: Interfaces with the ML-KEM IND-CCA2 challenger; forwards $(c_1^*, k_{1,b})$ as part of the challenge to $\mathcal{A}$; uses $\mathcal{O}_{\mathrm{MLKEM.Dec}}$ for oracle queries.
\[ |\Pr[\mathrm{Win}_{G_{1.5}}] - \Pr[\mathrm{Win}_{G_2}]| \leq \mathrm{Adv}^{\mathrm{IND\text{-}CCA2}}_{\mathrm{MLKEM}, B_{\mathrm{KEM1}}}. \]
\textbf{Game $G_3$ (DHKEM Randomization):} Replace $k_2^*$ with $k_{2,R} \leftarrow \{0,1\}^{256}$.
\emph{Reduction $B_{\mathrm{KEM2}}$}: Analogously interfaces with the DHKEM IND-CCA2 challenger.
\[ |\Pr[\mathrm{Win}_{G_2}] - \Pr[\mathrm{Win}_{G_3}]| \leq \mathrm{Adv}^{\mathrm{IND\text{-}CCA2}}_{\mathrm{DHKEM}, B_{\mathrm{KEM2}}}. \]
\textbf{Game $G_4$ (PRF Randomization):} With $k_1^*, k_2^*$ both uniform, $k_{\mathrm{sh}_0} = H(k_{1,R} \| k_{2,R} \| \cdots)$ is indistinguishable from random by PRF security of $H$.
\emph{Reduction $B_{\mathrm{PRF}}$}: Uses a PRF oracle to compute $k_{\mathrm{sh,chal}}$; outputs $\mathcal{A}$'s guess.
\[ |\Pr[\mathrm{Win}_{G_3}] - \Pr[\mathrm{Win}_{G_4}]| \leq \mathrm{Adv}^{\mathrm{PRF}}_{H, B_{\mathrm{PRF}}}. \]
\textbf{Final Bound:} $\Pr[\mathrm{Win}_{G_4}] = 1/2$. Summing:
\begin{align*}
\mathrm{Adv}^{\mathrm{IND\text{-}CCA2}}_{\mathrm{Auth},\mathcal{A}}
\;\leq\;&
2 \cdot \mathrm{Adv}^{\mathrm{EUF\text{-}CMA}}_{\mathrm{Rac}} \\
&+ \mathrm{Adv}^{\mathrm{IND\text{-}CCA2}}_{\mathrm{MLKEM}} \\
&+ \mathrm{Adv}^{\mathrm{IND\text{-}CCA2}}_{\mathrm{DHKEM}} \\
&+ \mathrm{Adv}^{\mathrm{PRF}}_{H}.
\end{align*}
All terms are negligible by assumption.
\renewcommand{\qedsymbol}{}
\end{proof}
\textbf{Non-Repudiation:} Both $\mathrm{sig}_s$ (binding server identity to its ephemeral KEM keys) and $\mathrm{sig}_c$ (binding client identity to the ciphertexts encapsulated to those keys) are publicly verifiable with their respective $k_{\mathrm{pub},\mathrm{rac}}$. EUF-CMA prevents forgery of either. Historical transcripts irrefutably link both parties to the full handshake exchange, even after offline key compromise (assuming pre-shared public keys are bound to identities).
\vspace{-3mm}
\subsection{Deniable Mode: Formal IND-CCA2 Game}
The ``ciphertext'' for the Deniable Mode AKEM is $(c, \mathrm{tag})$ and the session parameters include the server nonce $r_s$.
\textbf{AKEM-Den-IND-CCA2 Game:}
\begin{enumerate}[leftmargin=*]
    \item \textbf{Setup}: Challenger generates receiver ML-KEM and DHKEM keypairs. Challenger samples $r_s \leftarrow \{0,1\}^{128}$. $\mathcal{A}$ receives $(k_{\mathrm{pub}_{r1}}, k_{\mathrm{pub}_{r2}}, r_s)$.
    \item \textbf{Phase 1 (Decapsulation Oracle)}: $\mathcal{A}$ queries $O_{\mathrm{Decaps}}(c', \mathrm{tag}')$. Challenger: (i) decapsulates $(k_1', k_2')$; returns $\perp$ if either fails; (ii) derives $k_{\mathrm{auth}}'' = \mathrm{HKDF}(k_1' \| k_2' \| r_s, \text{``auth''})$; (iii) verifies $\mathrm{HMAC}(k_{\mathrm{auth}}'', c' \| r_s \| \texttt{MODE}) = \mathrm{tag}'$; returns $\perp$ if false; (iv) returns $k_{\mathrm{sh}}' = \mathrm{HKDF}(k_1' \| k_2' \| c' \| r_s \| k_{\mathrm{pub}_r} \| \texttt{MODE})$.
    \item \textbf{Challenge}: Challenger picks $b$; generates $(c^*, \mathrm{tag}^*, k_{\mathrm{sh}_0})$ per the protocol using $r_s$. If $b=0$: key $= k_{\mathrm{sh}_0}$; else: uniform. $\mathcal{A}$ receives $(c^*, \mathrm{tag}^*, k_{\mathrm{sh,chal}})$.
    \item \textbf{Phase 2}: $\mathcal{A}$ queries $O_{\mathrm{Decaps}}$ with $(c', \mathrm{tag}') \neq (c^*, \mathrm{tag}^*)$.
    \item \textbf{Guess}: $\mathcal{A}$ outputs $b'$; wins if $b' = b$.
\end{enumerate}
\vspace{-3mm}
\subsection{Deniable Mode: Sequence-of-Games Proof}
\begin{theorem}[Deniable Mode IND-CCA2]
\label{thm:den_conf}
If ML-KEM-768 and X25519-DHKEM are IND-CCA2 secure, HMAC-SHA-256 is MAC-Forge secure, and HKDF-SHA-256 is a secure PRF, then $k_{\mathrm{sh}}$ in the Deniable Mode is IND-CCA2 secure. For any PPT $\mathcal{A}$:
\[
\begin{aligned}
\mathrm{Adv}^{\mathrm{IND\text{-}CCA2}}_{\mathrm{Den},\mathcal{A}}
\;\leq\;&
\mathrm{Adv}^{\mathrm{MAC}}_{\mathrm{HMAC}} \\
&+ n_H \cdot \mathrm{Adv}^{\mathrm{PRF}}_{\mathrm{HKDF}} \\
&+ \mathrm{Adv}^{\mathrm{IND\text{-}CCA2}}_{\mathrm{MLKEM}} \\
&+ \mathrm{Adv}^{\mathrm{IND\text{-}CCA2}}_{\mathrm{DHKEM}}.
\end{aligned}
\]
where $n_H$ is the number of HKDF evaluations per session.
\end{theorem}
\begin{proof}
\textbf{Game $G_0$ (Real):} $\mathrm{Adv}_\mathcal{A}[G_0] = \mathrm{Adv}^{\mathrm{IND\text{-}CCA2}}_{\mathrm{Den},\mathcal{A}}$.\\
\textbf{Game $G_1$ (HMAC Unforgeability Abort):} Identical to $G_0$ except abort if $\mathcal{A}$ queries $O_{\mathrm{Decaps}}$ with $(c^*, \mathrm{tag}')$ where $\mathrm{tag}' \neq \mathrm{tag}^*$ and $\mathrm{HMAC}(k_{\mathrm{auth}}^*, c^* \| \texttt{MODE}) = \mathrm{tag}'$ (i.e., a valid tag forgery under the challenge authentication key).
\emph{Reduction $B_{\mathrm{MAC}}$}: Receives an HMAC oracle keyed with $k_{\mathrm{auth}}^*$. Simulates $G_0$, computing $k_{\mathrm{auth}}^*$ from $k_1^*, k_2^*$ via HKDF. If $\mathcal{A}$ produces such a forgery, $B_{\mathrm{MAC}}$ outputs it. The HKDF step introduces an additional $\mathrm{Adv}^{\mathrm{PRF}}_{\mathrm{HKDF}}$ term (bounding the probability that $k_{\mathrm{auth}}^*$ itself is distinguishable from random):
\[ |\Pr[\mathrm{Win}_{G_0}] - \Pr[\mathrm{Win}_{G_1}]| \leq \mathrm{Adv}^{\mathrm{MAC}}_{\mathrm{HMAC}} + \mathrm{Adv}^{\mathrm{PRF}}_{\mathrm{HKDF}}. \]
\textbf{Game $G_2$ (ML-KEM Randomization):} Replace $k_1^*$ with $k_{1,R} \leftarrow \{0,1\}^{256}$.
\emph{Reduction $B_{\mathrm{KEM1}}$}: Analogous to the Auth Mode proof; uses the ML-KEM IND-CCA2 challenger to embed the challenge. The HMAC oracle in $G_1$ is simulated by computing $k_{\mathrm{auth}}$ from the received $k_{1,b}$ (either $k_1^*$ or $k_{1,R}$). Thus:
\[ |\Pr[\mathrm{Win}_{G_1}] - \Pr[\mathrm{Win}_{G_2}]| \leq \mathrm{Adv}^{\mathrm{IND\text{-}CCA2}}_{\mathrm{MLKEM}, B_{\mathrm{KEM1}}}. \]
\textbf{Game $G_3$ (DHKEM Randomization):} Replace $k_2^*$ with $k_{2,R} \leftarrow \{0,1\}^{256}$.
\emph{Reduction $B_{\mathrm{KEM2}}$}: Analogous to $B_{\mathrm{KEM1}}$ for DHKEM. Thus:
\[ |\Pr[\mathrm{Win}_{G_2}] - \Pr[\mathrm{Win}_{G_3}]| \leq \mathrm{Adv}^{\mathrm{IND\text{-}CCA2}}_{\mathrm{DHKEM}, B_{\mathrm{KEM2}}}. \]
\textbf{Game $G_4$ (HKDF Randomization):} With $k_1^*, k_2^*$ both uniform, the HKDF derivations $n^*, k_{\mathrm{auth}}^*, k_{\mathrm{sh}_0}$ are all computationally indistinguishable from uniform random values by the PRF security of HKDF (applied $n_H$ times).
\emph{Reduction $B_{\mathrm{PRF}}$}: Uses a PRF oracle to compute the derived values; outputs $\mathcal{A}$'s guess. Accounts for $n_H$ HKDF invocations via a union bound:
\[ |\Pr[\mathrm{Win}_{G_3}] - \Pr[\mathrm{Win}_{G_4}]| \leq (n_H - 1) \cdot \mathrm{Adv}^{\mathrm{PRF}}_{\mathrm{HKDF}}. \]
In $G_4$, $k_{\mathrm{sh,chal}}$ is independent of $b$, so $\Pr[\mathrm{Win}_{G_4}] = 1/2$. Summing all transitions:
\[
\begin{aligned}
\mathrm{Adv}^{\mathrm{IND\text{-}CCA2}}_{\mathrm{Den},\mathcal{A}}
\;\leq\;&
\mathrm{Adv}^{\mathrm{MAC}}_{\mathrm{HMAC}} \\
&+ n_H \cdot \mathrm{Adv}^{\mathrm{PRF}}_{\mathrm{HKDF}} \\
&+ \mathrm{Adv}^{\mathrm{IND\text{-}CCA2}}_{\mathrm{MLKEM}} \\
&+ \mathrm{Adv}^{\mathrm{IND\text{-}CCA2}}_{\mathrm{DHKEM}}.
\end{aligned}
\]
All terms are negligible, completing the proof.\renewcommand{\qedsymbol}{}
\end{proof}
\vspace{-3mm}
\subsection{Sender Deniability: Formal Definition, Simulator, and Proof}
\begin{definition}[Offline Third-Party Sender Deniability]
\label{def:deniability}
A protocol is \emph{offline sender-deniable} if there exists a PPT simulator $Sim$ such that for all PPT adversaries $\mathcal{A}$ (``judges''):
\[
\begin{aligned}
\Big|
&\Pr\!\left[
    \mathcal{A}\!\left(
        T_{\mathrm{real}},\,
        k_{\mathrm{pub}_s},\,
        k_{\mathrm{pub}_r}
    \right)=1
\right] \\
&\quad -
\Pr\!\left[
    \mathcal{A}\!\left(
        T_{\mathrm{sim}},\,
        k_{\mathrm{pub}_s},\,
        k_{\mathrm{pub}_r}
    \right)=1
\right]
\Big|
\;\leq\; \mathrm{negl}(\lambda).
\end{aligned}
\]
where $T_{\mathrm{real}} = (c, \mathrm{tag})$ is generated by a real sender $S$ using $k_{\mathrm{pr}_s}$, and $T_{\mathrm{sim}} = (c', \mathrm{tag}')$ is generated by $Sim(k_{\mathrm{pr}_r})$ using only the receiver's private keys. The corruption model allows $\mathcal{A}$ to learn long-term secret keys of $S$ \emph{after} session completion (offline corruption), but ephemeral session keys are erased post-handshake and are unavailable. Note that this definition captures deniability toward third-party judges, not receiver-side entity authentication: the receiver accepts the session based on successful decapsulation and tag verification, not on a proof of sender identity.
\end{definition}
\begin{theorem}[Sender Deniability]
\label{thm:deniability}
The Deniable Mode is offline sender-deniable (Definition~\ref{def:deniability}). Given the server nonce $r_s$ (which the server generated itself), the simulator $Sim(k_{\mathrm{pr}_r}, r_s)$ described below achieves:
\[
\mathrm{Adv}_{\mathcal{A}} \leq \mathrm{Adv}^{\mathrm{IND\text{-}CCA2}}_{\mathrm{MLKEM}} + \mathrm{Adv}^{\mathrm{IND\text{-}CCA2}}_{\mathrm{DHKEM}} + \mathrm{Adv}^{\mathrm{PRF}}_{\mathrm{HKDF/HMAC}}.
\]
\end{theorem}
\begin{proof}
\vspace{-2mm}
\textbf{Simulator Construction.} $Sim(k_{\mathrm{pr}_r} = (k_{\mathrm{pr}_{r1}}, k_{\mathrm{pr}_{r2}}), r_s)$:
\begin{enumerate}[leftmargin=*]
    \item Generate fresh ciphertexts targeting the receiver's own public keys: $(c_1, k_1) \leftarrow \mathrm{MLKEM.Enc}(k_{\mathrm{pub}_{r1}})$, $(c_2, k_2) \leftarrow \mathrm{DHKEM.Enc}(k_{\mathrm{pub}_{r2}})$.
    \item Compute $k_1 \leftarrow \mathrm{MLKEM.Dec}(k_{\mathrm{pr}_{r1}}, c_1)$, $k_2 \leftarrow \mathrm{DHKEM.Dec}(k_{\mathrm{pr}_{r2}}, c_2)$.
    \item Derive $k_{\mathrm{auth}} \leftarrow \mathrm{HKDF}(k_1 \| k_2 \| r_s, \text{``auth''})$.
    \item Compute $\mathrm{tag} \leftarrow \mathrm{HMAC}(k_{\mathrm{auth}}, c \| r_s \| \texttt{MODE})$ where $c = c_1 \| c_2$.
    \item Output $T_{\mathrm{sim}} = (c, \mathrm{tag})$.
\end{enumerate}
\textbf{Why the simulator has access to $r_s$:} The server nonce $r_s$ is transmitted in the Server Hello and is known to the receiver (who generated it), the sender (who received it in plaintext), and the simulator (who is the receiver). Using $r_s$ in the simulator is legitimate: the simulator acts as the receiver playing the role of a fake sender. Since $r_s$ is public within the session, the simulator possessing it does not reduce deniability where a judge also observes $r_s$ in the transcript.\\
\textbf{Perfect Simulation:} Both the sender $S$ and $Sim$ derive $(k_1, k_2)$ from the same ciphertexts $(c_1, c_2)$ under the receiver's public keys: $S$ via encapsulation and $Sim$ via decapsulation of its own freshly generated ciphertexts. By KEM correctness, $\mathrm{MLKEM.Dec}(k_{\mathrm{pr}_{r1}}, c_1) = k_1$ and $\mathrm{DHKEM.Dec}(k_{\mathrm{pr}_{r2}}, c_2) = k_2$. All downstream derivations ($k_{\mathrm{auth}}$, $\mathrm{tag}$) are identical deterministic functions of the same inputs $(k_1, k_2, r_s, c, \texttt{MODE})$. For the same $(c_1, c_2)$, $T_{\mathrm{sim}} = T_{\mathrm{real}}$ exactly.
\textbf{Indistinguishability Argument:} When $Sim$ generates independent fresh ciphertexts, the joint distribution $(c^{Sim}, \mathrm{tag}^{Sim})$ differs statistically from $(c^S, \mathrm{tag}^S)$ only in the ciphertexts. By IND-CCA2 of ML-KEM and DHKEM, no PPT $\mathcal{A}$ can distinguish the ciphertext distributions; by PRF security of HKDF and HMAC, the derived $k_{\mathrm{auth}}$ and tag values are indistinguishable from random given the (indistinguishable) KEM secrets. Hence:
\[
\begin{aligned}
\Big|
\Pr\!\left[\mathcal{A}(T_{\mathrm{real}})=1\right]
&- \Pr\!\left[\mathcal{A}(T_{\mathrm{sim}})=1\right]
\Big| \\
&\leq \mathrm{Adv}^{\mathrm{IND\text{-}CCA2}}_{\mathrm{MLKEM}} \\
&\quad + \mathrm{Adv}^{\mathrm{IND\text{-}CCA2}}_{\mathrm{DHKEM}} \\
&\quad + \mathrm{Adv}^{\mathrm{PRF}}_{\mathrm{HKDF/HMAC}}.
\end{aligned}
\]
\textbf{Offline Corruption Resilience:} After session completion, all ephemeral keys are erased. The HMAC tag binds to the ephemeral KEM secrets $(k_1, k_2)$ and $r_s$---none of which are derivable from the sender's long-term identity key alone. Corrupting the sender's long-term key post-session reveals no information distinguishing a real transcript from a receiver-fabricated one. All terms negligible.
\renewcommand{\qedsymbol}{}
\end{proof}
\vspace{-5mm}
\section{Experimental Setup}
\label{sec:setup}
\subsection{Hardware and Testbed Configuration}
The evaluation testbed simulated the heterogeneous hardware landscape of a deployed smart grid. The \textbf{server node} (representing a utility SCADA control center or regional distribution management system) ran on a virtual machine provisioned with 4 dedicated CPU cores of an AMD Ryzen 5 8640HS processor (Zen 4 architecture, x86-64, 3.5\,GHz base clock, 4.9\,GHz boost, shared L3 cache). The \textbf{laptop client node} used the same AMD Ryzen 5 architecture, representing a medium-capability field engineer terminal. The \textbf{Raspberry Pi 500 client node} (ARM Cortex-A76 cluster, 2.4\,GHz, 8\,GB LPDDR4X RAM, equivalent SoC to Raspberry Pi 5) represented a resource-constrained smart meter or substation edge device---the primary performance-bottleneck target of this evaluation. The server--Pi topology models the production deployment path: a high-capacity SCADA system communicating with a low-power smart meter over a WAN or LPWAN-bridged link.
\vspace{-3mm}
\subsection{Software Stack}
Implementations were written in C with the following libraries: \textbf{liboqs~0.10.1} (Open Quantum Safe) for ML-KEM-768 (Kyber reference implementation with AVX2 optimization enabled on x86 targets); \textbf{libsodium~1.0.19} for X25519-DHKEM in HPKE mode per RFC~9180 (constant-time Montgomery ladder implementation); \textbf{Raccoon reference C implementation}~\cite{delpino2024raccoon} (first-order masked, SHAKE-based); \textbf{ML-DSA-65} (FIPS~204 reference via liboqs, included for comparison only); \textbf{Ascon-128 reference C implementation} for AEAD. Compiler: GCC~13.2 with \texttt{-O2} and no auto-vectorization flags beyond platform defaults. All benchmarks are wall-clock averages over 1,000 independent iterations with standard deviation verified to be $<$5\% of the mean. Communication used HTTP/1.1 over TCP for prototyping; the protocol is transport-agnostic and targets ANSI C12.22/TLS~1.3 deployment in production.
\vspace{-3mm}
\subsection{Key Management Model}
ML-KEM-768 and X25519-DHKEM keypairs were generated ephemerally at the start of each session. For isolated cryptographic benchmarking, Raccoon keypairs were also generated ephemerally and measured independently (production deployments use pre-shared long-term Raccoon keys, avoiding public key transmission over the network).
\vspace{-3mm}
\subsection{Evaluation Scenarios and Metrics}
Three evaluation scenarios were executed: (1) \emph{Authenticated Communication}: legitimate clients completed the full Auth Mode handshake and exchanged ANSI C12.18 table data encrypted with Ascon-128; (2) \emph{Authentication Failure Handling}: clients with invalid credentials were rejected, with authentication failure events logged and verified not to expose timing oracles; (3) \emph{Network Delay Resilience}: manually introduced 10--100\,ms one-way delays to verify the protocol completes correctly under WAN conditions. Performance metrics collected: per-operation time (ns) and CPU cycle count via RDTSC; end-to-end handshake latency (sender + network + receiver); derived throughput ($\text{Throughput} = (1500 \times 8) / \text{latency}$ Mbps, per ANSI C12.22 MTU); total handshake payload size (bytes); and CPU cycle total as an energy consumption proxy.
\vspace{-3mm}
\section{Results and Discussion}
\label{sec:results}

\subsection{Key Generation Performance}
Table~\ref{tab:keygen_combined} reports per-operation key generation costs averaged over 1,000 iterations. Raccoon keypair generation is the most expensive individual operation ($\sim$2.4M cycles on Ryzen~5, $\sim$6M cycles on Pi), attributable to the masking structure's $O(d \log d)$ overhead. However, this cost is amortized over the session lifetime in production: long-term Raccoon keys are provisioned during device manufacturing or registration and reused across all sessions. Per-session ephemeral key generation (ML-KEM-768 and X25519) is fast on all platforms: ML-KEM takes $\sim$148\,K cycles (server) to $\sim$319\,K cycles (Pi); X25519 takes $\sim$55\,K cycles (server) to $\sim$249\,K cycles (Pi). The Pi's ARM architecture shows a consistent $\sim$2.4$\times$ penalty over Ryzen~5 for these operations.

\begin{table}[htbp]
\centering
\caption{Average Performance of Key Generation Operations (1,000 iterations)}
\label{tab:keygen_combined}
\begin{tabular}{@{}l l c c@{}}
\toprule
\textbf{Scheme} & \textbf{Participant} & {\textbf{Time (ns)}} & {\textbf{Cycles}} \\
\midrule
\multirow{3}{*}{X25519} & Server & 15767 & 54995 \\
 & Ryzen 5 & 13993 & 48792 \\
 & Pi & 103685 & 248844 \\
\addlinespace
\multirow{3}{*}{ML-KEM-768} & Server & 42293 & 147657 \\
 & Ryzen 5 & 49631 & 173266 \\
 & Pi & 133074 & 319377 \\
\addlinespace
\multirow{3}{*}{Raccoon DSA} & Server & 737109 & 2574178 \\
 & Ryzen 5 & 678511 & 2370045 \\
 & Pi & 2489383 & 5974519 \\
\addlinespace
\multirow{2}{*}{ML-DSA-65} & Server & 124145 & 433567 \\
 & Ryzen 5 & 139276 & 486290 \\
\bottomrule
\end{tabular}
\end{table}
\vspace{-3mm}
\subsection{Authenticated Mode: Per-Operation Costs}

\subsubsection{Sender (Server) Operations}
Table~\ref{tab:auth_sender} reports the server-side per-operation costs. Raccoon signing (1,225\,$\mu$s server-side) is the dominant server cost, but the server initiates by signing its identity assertion---this cost is amortized over the session lifetime, not per-packet. ML-KEM-768 encapsulation (56\,$\mu$s) and X25519-DHKEM encapsulation (39\,$\mu$s) are sub-millisecond and contribute negligibly to total handshake time.

\begin{table}[h!]
\centering
\footnotesize
\caption{Authenticated Mode: Sender (Server) Operation Costs, AMD Ryzen 5}
\label{tab:auth_sender}
\begin{tabular}{@{}lcc@{}}
\toprule
\textbf{Operation} & {\textbf{Time (ns)}} & {\textbf{Cycles}} \\
\midrule
ML-KEM-768 Encapsulation   &   56196 &   196157 \\
X25519-DHKEM Encapsulation  &   39041 &   136132 \\
Raccoon Signing             & 1225127 &  4279419 \\
ML-DSA-65 Signing (baseline)&  489609 &  1710203 \\
Raccoon Verification        &  408700 &  1430500 \\
Ascon-128 AEAD Encryption   &    2686 &     9401 \\
\bottomrule
\end{tabular}
\end{table}

\subsubsection{Receiver (Client) Operation Costs}
Table~\ref{tab:auth_receiver_updated} details the client-side receiver costs for both Ryzen~5 and Pi platforms. On the Pi, Raccoon signing (4,556\,$\mu$s, $\sim$16M cycles) is the overall bottleneck, nearly $3.7\times$ slower than on Ryzen~5 due to architectural differences between ARM and x86 in the NTT operations. Raccoon verification (408.7\,$\mu$s Ryzen~5, 727.3\,$\mu$s Pi) is $\sim$7$\times$ costlier than ML-KEM decapsulation alone, and dominates the server's total receiver-side latency ($\sim$840\,$\mu$s Ryzen~5, $\sim$1.3\,ms Pi).

\begin{table*}[htbp]
\centering
\footnotesize
\caption{Authenticated Mode: Client-Side Operation Costs on Ryzen 5 vs.\ Raspberry Pi 500}
\label{tab:auth_receiver_updated}
\begin{tabular}{@{}l S[table-format=6.1] S[table-format=6.1] S[table-format=6.1] S[table-format=6.1]@{}}
\toprule
 & \multicolumn{2}{c}{\textbf{Ryzen 5 Client}} & \multicolumn{2}{c}{\textbf{Pi 500 Client}} \\
\cmidrule(lr){2-3} \cmidrule(lr){4-5}
\textbf{Operation} & {\textbf{Time ($\mu$s)}} & {\textbf{Cycles (k)}} & {\textbf{Time ($\mu$s)}} & {\textbf{Cycles (k)}} \\
\midrule
Raccoon Sign               & 1225.1 &  4279.4 & 4556.3 & 15945.9 \\
Raccoon Signature Verify   &  408.7 &  1430.5 &  727.3 &  2545.4 \\
ML-DSA-65 Sign (baseline)  &  489.6 &  1710.2 & N/A    & N/A \\
ML-DSA-65 Verify (baseline)&  121.7 &   426.0 &  257.8 &   924.2 \\
ML-KEM-768 Decapsulation   &   59.1 &   206.9 &   74.1 &   259.3 \\
X25519-DHKEM Decapsulation  &   33.3 &   116.4 &   38.9 &   136.3 \\
Ascon-128 AEAD Decrypt     &    6.0 &    20.9 &    1.7 &     5.8 \\
\bottomrule
\end{tabular}
\end{table*}

\subsubsection{Throughput and End-to-End Latency}
\textbf{Throughput methodology.} The throughput figures represent the \emph{maximum session-establishment rate} for 1,500-byte MTU payloads under ANSI C12.22, computed as $\text{Throughput} = (1500 \times 8\,\text{bits}) / T_{\mathrm{handshake}}$ Mbps, where $T_{\mathrm{handshake}}$ is the \emph{one-way, single-operation time} measured at the receiver for the dominant handshake operation. It is \emph{not} a per-packet data rate: post-handshake data is encrypted with Ascon-128 at negligible additional cost.

For the Deniable Mode, the dominant receiver cost is decapsulation ($\sim$87\,$\mu$s Ryzen~5, $\sim$409\,$\mu$s Pi), giving throughput of $(12000\,\text{bits}) / 87\,\mu\text{s} \approx 138$\,Mbps at the Ryzen~5 receiver. For the Authenticated Mode with Raccoon, the dominant receiver cost is signature verification ($\sim$507\,$\mu$s Ryzen~5 total receiver side), giving $(12000\,\text{bits}) / 507\,\mu\text{s} \approx 23.7$\,Mbps per decapsulation; the full round-trip handshake (including Pi signing at 4.8\,ms) gives $\sim$1.6\,Mbps per end-to-end session. Table~\ref{tab:throughput} reports all scenarios with explicit methodology.

\begin{table}[h!]
\centering
\caption{Derived Session-Establishment Throughput (1,500-byte MTU)}
\label{tab:throughput}
\begin{tabular}{@{}lp{3.5cm}S[table-format=3.1]@{}}
\toprule
\textbf{Scenario} & \textbf{Description} & {\textbf{Throughput (Mbps)}} \\
\midrule
A   & Deniable Mode (Ryzen~5 receiver, one-way)   & 138.0 \\
B-1 & Auth Mode, Raccoon (Ryzen~5 receiver, one-way) &  23.7 \\
B-2 & Auth Mode, Raccoon (Pi full round-trip)     &   1.6 \\
B-3 & Auth Mode, ML-DSA-65 (Ryzen~5, one-way)    &  52.2 \\
\bottomrule
\end{tabular}
\captionsetup{font=small}
\caption*{\textbf{Methodology:} Throughput $= (1500 \times 8\,\text{bits}) / T$ Mbps. One-way: $T$ = receiver-side dominant operation. Full round-trip (B-2): $T \approx 7.3$\,ms including Pi signing (4.8\,ms) + server verification (0.84\,ms) + network. This is a session-establishment rate, not per-packet data rate; post-handshake encryption via Ascon-128 adds $<$3\,$\mu$s per payload.}
\vspace{-6mm}
\end{table}
\vspace{-3mm}
\subsection{Deniable Mode Performance}
Table~\ref{tab:deniable_updated} presents detailed Deniable Mode per-operation costs. The HMAC tag computation (4.0\,$\mu$s server, 4.6\,$\mu$s client) and HKDF nonce/key derivation (55\,$\mu$s server, 167\,$\mu$s client) contribute minimally compared to the KEM operations. Total sender latency: 154\,$\mu$s (0.15\,ms server); total receiver latency: 409\,$\mu$s (0.41\,ms Pi). This $5.5\times$ improvement over the Auth Mode server receiver latency (840\,$\mu$s) is entirely attributable to eliminating Raccoon signature verification, confirming that the symmetric HMAC approach is highly efficient for deniable authentication.

\begin{table}[h!]
\centering
\caption{Deniable Mode: Per-Operation Costs on Server and Pi 500}
\label{tab:deniable_updated}
\begin{tabular}{@{}lccc@{}}
\toprule
\textbf{Operation} & \textbf{Role} & {\textbf{Time (ns)}} & {\textbf{Cycles}} \\
\midrule
\multirow{2}{*}{X25519-DHKEM Enc/Dec} & Server &  39041 & 136132 \\
 & Pi Client & 103685 & 248844 \\
\addlinespace
\multirow{2}{*}{ML-KEM-768 Enc/Dec} & Server &  56196 & 196157 \\
 & Pi Client & 133074 & 319377 \\
\addlinespace
\multirow{2}{*}{HKDF (nonce + key)} & Server &  55185 & 193147 \\
 & Pi Client & 167481 & 401954 \\
\addlinespace
\multirow{2}{*}{HMAC Tag} & Server &   4008 &  13918 \\
 & Pi Client &   4617 &  11080 \\
\bottomrule
\end{tabular}

\end{table}
\vspace{-3mm}
\subsection{Payload Sizes and Network Footprint}
Table~\ref{tab:data_sizes_updated} presents a full breakdown of cryptographic component and handshake payload sizes. The 12,644-byte Authenticated Mode payload (with Raccoon) is dominated by the 11,524-byte signature---a well-known characteristic of Raccoon's security-performance trade-off. This payload may require fragmentation across ANSI C12.22 APDUs and is unsuitable for LoRaWAN or Sigfox-based LPWAN links (typical MTU: 51--256 bytes). However, for wired or WiFi-connected substation controllers, 12,644 bytes is transmitted in $<$1\,ms on a 100\,Mbps link---acceptable for infrequent control session initiation. The 4,413-byte Auth Mode payload with ML-DSA-65 reduces this overhead substantially but eliminates SCA resistance. The 1,152-byte Deniable Mode payload is LPWAN-compatible after APDU fragmentation and is the preferred mode for telemetry-heavy deployments.

\begin{table}[t]
\centering
\caption{Cryptographic Component and Handshake Payload Sizes (Bytes)}
\label{tab:data_sizes_updated}
\renewcommand{\arraystretch}{1.1}  
\begin{tabular}{@{}lcc@{}}
\toprule
\textbf{Component}                  & \textbf{Pub. Key (B)} & \textbf{C/S/T (B)} \\
\midrule
X25519-DHKEM                        & 32                    & 32                 \\
ML-KEM-768                          & 1184                  & 1088               \\
ML-DSA-65                           & 1952                  & 3293               \\
Raccoon DSA                         & 2256                  & 11524              \\
HMAC-SHA-256 Tag                    & --                    & 32                 \\
\midrule
\textbf{Auth Mode (Raccoon)}        & \textbf{3472}         & \textbf{12644}     \\
\textbf{Auth Mode (ML-DSA-65)}      & \textbf{3168}         & \textbf{4413}      \\
\textbf{Deniable Mode}              & \textbf{1216}         & \textbf{1152}      \\
\bottomrule
\end{tabular}
\end{table}
\vspace{-3mm}
\subsection{Resource Utilization and Energy Efficiency}
Table~\ref{tab:performance_combined} consolidates CPU cycle counts across all operations and both modes. The Deniable Mode requires only $\sim$981\,K cycles total on the Pi (0.41\,ms), compared to $\sim$18.5M cycles for a full Authenticated Mode handshake (Raccoon sign + verify on Pi: $\sim$15.9M + $\sim$2.5M cycles). This 93\% reduction in CPU cycles directly translates to lower energy consumption---critical for battery-powered or energy-harvesting smart meter designs. For the Ryzen~5 server, the Deniable Mode receiver costs 87\,$\mu$s vs.\ 840\,$\mu$s for Auth Mode verification, a $9.7\times$ speedup enabling proportionally higher session concurrency.

\begin{table}[htbp]
\centering
\caption{Performance Summary Across Platforms and Modes}
\label{tab:performance_combined}
\resizebox{\columnwidth}{!}{%
\begin{tabular}{@{}l c c c@{}}
\toprule
\textbf{Operation} & \textbf{Platform} & {\textbf{Time ($\mu$s)}} & {\textbf{Cycles (k)}} \\
\midrule
\multicolumn{4}{l}{\textit{Key Generation (per-session ephemeral)}} \\
\multirow{3}{*}{X25519}     & Server   &  15.8 &   55.0 \\
                             & Ryzen 5  &  14.0 &   48.8 \\
                             & Pi       & 103.7 &  248.8 \\
\multirow{3}{*}{ML-KEM-768} & Server   &  42.3 &  147.7 \\
                             & Ryzen 5  &  49.6 &  173.3 \\
                             & Pi       & 133.1 &  319.4 \\
\multirow{3}{*}{Raccoon DSA (long-term)} & Server &  737.1 & 2574.2 \\
                             & Ryzen 5  & 678.5 & 2370.0 \\
                             & Pi       & 2489.4 & 5974.5 \\
\midrule
\multicolumn{4}{l}{\textit{Authenticated Mode Handshake}} \\
\multirow{2}{*}{Sender Sign}  & Ryzen 5 & 1225.1 &  4279.4 \\
                               & Pi      & 4556.3 & 15945.9 \\
\multirow{2}{*}{Receiver Verify} & Ryzen 5 &  408.7 & 1430.5 \\
                                  & Pi      &  727.3 & 2545.4 \\
\midrule
\multicolumn{4}{l}{\textit{Deniable Mode Handshake (total)}} \\
Sender total  & Server   & 154.4 &  539.4 \\
\multirow{2}{*}{Receiver total} & Ryzen 5 &  87.0 &  303.3 \\
                                 & Pi      & 408.9 &  981.3 \\
\bottomrule
\end{tabular}%
}\vspace{-4mm}
\end{table}
\vspace{-3mm}
\subsection{State-of-the-Art Performance Comparison}
Table~\ref{tab:perf_security_comparison} situates AmphiKey within the broader landscape. AmphiKey's Deniable Mode matches hybrid HPKE in payload (1,152 bytes) and latency ($\sim$0.41\,ms Pi) while adding integrated deniable HMAC authentication which is a zero-overhead addition for functionality. AmphiKey's Deniable Mode outperforms Shadowfax~\cite{cryptoeprint:2025/154} by 35\% in payload size and provides cleaner, PRF-based deniability proofs. The Authenticated Mode is the only protocol in the comparison providing all three of: PQ confidentiality, non-repudiable authentication, and SCA-resistant signing---at the cost of a larger payload required by Raccoon's security architecture.

\begin{table*}[htbp]
\centering
\footnotesize
\caption{Performance and Security Comparison Against State-of-the-Art}
\label{tab:perf_security_comparison}
\begin{tabular}{@{}l p{3.8cm} c c c c@{}}
\toprule
\textbf{Protocol} & \textbf{Core Primitives} & \textbf{Formal} & \textbf{Non-Rep.} & \textbf{Payload} & \textbf{Client Latency} \\
 & & \textbf{Deniability} & \textbf{Auth.} & \textbf{(Bytes)} & \textbf{(Raspberry Pi)} \\
\midrule
Hybrid HPKE~\cite{rfc9180}             & X25519, ML-KEM-768, AEAD         & \texttimes & \texttimes & 1,152  & $\sim$0.41\,ms \\
X-Wing~\cite{xwing_kem}                & X25519, ML-KEM-768, SHA3          & \texttimes & \texttimes & 1,120  & $<$0.41\,ms \\
Shadowfax~\cite{cryptoeprint:2025/154} & NIKE + PQ KEM + PQ Ring Sig       & \checkmark & \texttimes & 1,781  & $\sim$0.48\,ms \\
Signal PQXDH~\cite{signal_pqxdh}       & X3DH + ML-KEM-768                 & Informal   & \texttimes & N/A    & N/A \\
\midrule
\textbf{AmphiKey Auth (Raccoon)}   & Hybrid KEM + Raccoon DSA (masked) & \texttimes & \checkmark & \textbf{12,644} & \textbf{$\sim$4.8\,ms} \\
\textbf{AmphiKey Deniable}         & Hybrid KEM + HMAC-SHA-256         & \checkmark & \texttimes & \textbf{1,152}  & \textbf{$\sim$0.41\,ms} \\
\bottomrule
\end{tabular}
\captionsetup{font=small}

\end{table*}
\vspace{-3mm}
\subsection{Smart Grid Deployment Applicability}
The dual-mode design maps directly to ANSI C12.22 traffic class requirements. Authenticated Mode suits infrequent, safety-critical control sessions: SCADA relay commands, firmware updates, and substation configuration changes occur at $<$1 session/hour on typical distribution feeders, making 4.8\,ms Pi signing latency and 12,644-byte payloads acceptable. Deniable Mode is optimal for high-frequency telemetry: smart meters reporting at 15-minute intervals and PMU phasor data streams all benefit from 0.41\,ms latency and 1,152-byte payloads that fit comfortably within IEEE C37.118.2 and ANSI C12.22 APDU constraints. A mass-reconnect scenario following a grid outage (thousands of meters attempting simultaneous re-authentication) represents the critical dimensioning case: in Authenticated Mode, each Pi takes 4.8\,ms to sign, limiting concurrency to $\sim$208 sessions/second per Pi; in Deniable Mode, 0.41\,ms per session allows $\sim$2,439 sessions/second---a $11.7\times$ improvement that could be the difference between orderly reconnection and a cascading control failure.
\section{Conclusion}
\label{sec:conclusion}
AmphiKey is the first dual-mode PQ/classical hybrid AKEM to jointly address algorithmic, quantum, and physical attack vectors for smart grid communications, with complete formal security proofs. The Deniable Mode (0.41\,ms, 1,152 bytes, Raspberry Pi) delivers efficient privacy-preserving security with proven sender deniability, replay resistance via ephemeral nonce derivation, and 93\% lower computational cost than the Authenticated Mode. The Authenticated Mode (4.8\,ms Pi signing, 12,644 bytes) provides non-repudiable, SCA-resistant authentication anchored by Raccoon DSA's architectural masking support, with explicit characterization of SCA scope: Raccoon signing is protected; ML-KEM and X25519 operations require additional deployment-level countermeasures. Both modes incorporate cryptographic \texttt{MODE} binding preventing active downgrade attacks. The security of both modes is proven via complete sequence-of-games reductions: Theorems~\ref{thm:auth_conf}--\ref{thm:deniability} formally ground IND-CCA2 and offline sender deniability in standard EUF-CMA, IND-CCA2, and PRF assumptions. Future work includes TLS~1.3 integration via a hybrid extension, exploration of higher-order Raccoon masking for ARM targets, hardware-accelerated KEM implementations for sub-100\,$\mu$s Pi handshakes, and pilot deployment on a live ANSI C12.22 grid segment.
\vspace{-3mm}
\section*{Acknowledgment}
This material is based on work supported by the U.S. Department of Energy, Office of Cybersecurity, Energy Security, and Emergency Response (CESER), under Award Number DE-CR0000050. This report was prepared as an account of work sponsored by an agency of the United States Government. Neither the United States Government nor any agency thereof, nor any of their employees, makes any warranty, express or implied, or assumes any legal liability or responsibility for the accuracy, completeness, or usefulness of any information, apparatus, product, or process disclosed, or represents that its use would not infringe privately owned rights. The views and opinions of authors expressed herein do not necessarily state or reflect those of the United States Government or any agency thereof.

\end{document}